\documentclass[twocolumn]{openjournal}

\usepackage{bm}
\usepackage{bold-extra}
\usepackage{xspace}
\usepackage{xcolor}
\usepackage{amsmath}
\usepackage{graphicx}
\usepackage[hidelinks]{hyperref}
\hypersetup{breaklinks=true,colorlinks=true,linkcolor=blue,urlcolor=blue,citecolor=blue}
\usepackage[caption=false]{subfig}
\usepackage{booktabs}
\usepackage[capitalise]{cleveref}

\usepackage{savesym}
\savesymbol{tablenum}
\usepackage{siunitx}
\restoresymbol{SIX}{tablenum}

\usepackage{tabularx}

\usepackage{lineno}

\usepackage{ragged2e}

\newcommand{\hii}{\relax \ifmmode {\mbox H\,{\scshape ii}}\else H\,{\scshape ii}\fi}
\newcommand{\mi}{\relax \ifmmode {\mu{\mbox m}}\else $\mu$m\fi}
\newcommand{\ha}{\relax \ifmmode {\mbox H}\alpha\else H$\alpha$\fi}
\newcommand{\hb}{\relax \ifmmode {\mbox H}\beta\else H$\beta$\fi}

\newcommand{\sii}{\relax \ifmmode {\mbox S\,{\scshape ii}}\else S\,{\scshape ii}\fi}
\newcommand{\siii}{\relax \ifmmode {\mbox S\,{\scshape iii}}\else S\,{\scshape iii}\fi}
\newcommand{\siv}{\relax \ifmmode {\mbox S\,{\scshape iv}}\else S\,{\scshape iv}\fi}
\newcommand{\nii}{\relax \ifmmode {\mbox N\,{\scshape ii}}\else N\,{\scshape ii}\fi}
\newcommand{\oi}{\relax \ifmmode {\mbox O\,{\scshape i}}\else O\,{\scshape i}\fi}
\newcommand{\oii}{\relax \ifmmode {\mbox O\,{\scshape ii}}\else O\,{\scshape ii}\fi}
\newcommand{\oiii}{\relax \ifmmode {\mbox O\,{\scshape iii}}\else O\,{\scshape iii}\fi}
\newcommand{\neiii}{\relax \ifmmode {\mbox Ne\,{\scshape iii}}\else Ne\,{\scshape iii}\fi}

\newcommand{\rdostres}{\relax \ifmmode {\,\mbox{R}}_{\rm 23}\else \,\mbox{R}$_{\rm 23}$\fi} 

\newcommand{\ciii}{\relax \ifmmode {\mbox O\,{\scshape iii}}\else C\,{\scshape iii}\fi}
\newcommand{\civ}{\relax \ifmmode {\mbox O\,{\scshape iii}}\else C\,{\scshape iv}\fi}
\newcommand{\nv}{\relax \ifmmode {\mbox N\,{\scshape v}}\else N\,{\scshape v}\fi}
\newcommand{\heii}{\relax \ifmmode {\mbox He\,{\scshape ii}}\else He\,{\scshape ii}\fi}

\newcommand{\gsim}{\hbox{\rlap{\lower.55ex\hbox{$\sim$}} \kern-.3em
\raise.4ex \hbox{$>$}}}
\newcommand{\lsim}{\hbox{\rlap{\lower.55ex\hbox{$\sim$}} \kern-.3em
\raise.4ex \hbox{$<$}}}

\shorttitle{Sulphur abundances in AGN with HCm}
\shortauthors{P\'erez-Montero et al.}

\begin{document}

\journalinfo{The Open Journal of Astrophysics}

\title{HII-CHI-mistry for active galactic nuclei: optical sulphur abundances without empirical ionisation correction factors}

\author{E. P\'erez-Montero$^{1,\ast}$, O.~L. Dors Jr.$^2$, I.~A. Zinchenko$^{3,4}$, B. P\'erez-D\'{\i}az$^{5,1}$, J.~M. V\'{\i}lchez$^1$ }
\email{$^\ast$epm@iaa.es}

\affiliation{
$^1$Instituto de Astrof\'\i sica de Andaluc\'\i a. CSIC. Apartado de correos 3004. 18080, Granada, Spain\\
$^2$Universidade do Vale do Para\'iba, Av. Shishima Hifumi, 2911, Cep
12244-000, S\~ao Jos\'e dos Campos, SP, Brazil\\
$^3$ Astronomisches Rechen-Institut, Zentrum f\"{u}r Astronomie der Universit\"{a}t Heidelberg, M\"{o}nchhofstra{\ss}e 12-14, D-69120 Heidelberg, Germany \\
$^4$Main Astronomical Observatory, National Academy of Sciences of Ukraine, 27 Akad. Zabolotnoho St 03680 Kyiv, Ukraine\\
$^5$INAF - Osservatorio Astronomico di Roma, via Frascati 33, I-00078, Monte Porzio Catone, Italy\\
}

\begin{abstract}
The determination of sulphur abundances in active galactic nuclei (AGN) remains uncertain because the contribution of unobserved ionisation stages is commonly estimated using ionisation correction factors (ICFs) calibrated for \hii\ regions. In this work, we extend the {\sc HII-CHI-mistry} methodology to derive total sulphur abundances in AGN from optical emission lines through comparison with photoionisation models. The new implementation simultaneously provides S/H using model grids covering a range of sulphur abundances and ionising spectral energy distributions (SEDs).
The method is tested using a control sample of Seyfert galaxies with ionic abundances derived through the direct method. The best agreement between {\sc HCm} and the observed ionic abundances was obtained for AGN models with $\alpha_{\rm ox}=-1.2$ and a stopping criterion corresponding to a free-electron fraction of 98\%. While oxygen abundances are consistent with previous determinations, sulphur abundances are systematically larger than those obtained from direct-method analyses using classical ICFs, suggesting that these prescriptions underestimate the contribution of higher ionisation sulphur stages in AGN.
We applied the method to a large sample of  Seyfert and LINER-like regions from the MaNGA survey,
with all populations preserving a common positive S/O-metallicity relation.
Moreover, although with a larger intrinsic scatter, Seyfert regions tend to display slightly larger median S/O ratios than star-forming regions at high oxygen abundances, while LINERs occupy an intermediate position between both populations. When S/H is adopted as the metallicity tracer instead of O/H, all populations still define similar increasing S/O sequences. The modest differences between star-forming regions and Seyferts become even weaker, although Seyferts still tend to occupy the upper envelope of the relation, suggesting that sulphur may provide a more robust tracer of the total gas-phase metallicity while leaving room for additional AGN-related effects. Since all AGN regions were analysed using the same photoionisation framework, the observed differences between Seyferts and LINERs may reflect intrinsic differences in their ionisation conditions and possibly in their ionising SEDs. These results indicate that sulphur abundances provide additional constraints on both the chemical properties and the ionisation structure of AGN narrow-line regions.
\end{abstract}

\maketitle



\section{Introduction}\label{sec:intro}

The gas-phase chemical composition of the narrow-line regions (NLRs) of
active galactic nuclei (AGN) provides a fundamental tool to investigate
the interplay between chemical enrichment, star formation history, gas
accretion, feedback processes and black-hole growth in galaxy nuclei \citep{nagao06,dors14,flury20}.
Optical emission-line spectroscopy has long enabled the determination of the physical conditions and chemical abundances of ionised nebulae.
However, abundance determinations in AGN
remain substantially more uncertain than those in classical \hii\ regions.
This is mainly because the harder ionising radiation fields in AGN
produce gas with a high degree of ionisation and, consequently,
significant fractions of high-ionisation species,
such as $\rm O^{3+}$ and $\rm S^{3+}$. Since the emission lines of these ions are
not observed in the optical spectrum, their contributions can be difficult to constrain, even though they may represent a significant fraction of the total elemental abundance budget.
In addition, AGN NLRs may present complex and poorly constrained density structures (e.g. \citealt{zhang13,Rodriguez2017,binette22}). Moreover, shocks can contribute to the observed emission-line fluxes and modify the thermal and ionisation structure of the emitting gas (e.g. \citealt{dors21}). These effects introduce additional uncertainties in abundance determinations and further complicate the use of prescriptions calibrated for classical H ii regions.

Among the alpha-elements, sulphur constitutes a particularly interesting tracer. Sulphur is mainly produced by massive stars and broadly follows oxygen enrichment \citep{garnett97,kehrig06}, but unlike oxygen it is expected to suffer weaker depletion onto dust grains (e.g. \citealt{garnett89,savage96}). The sulphur-to-oxygen abundance ratio therefore provides relevant complementary information on chemical evolution under the dust processing conditions  of dusty nuclear environments.

The code {\sc HII-CHI-mistry} (\citealt{hcm14}, hereafter {\sc HCm})\footnote{All the different versions of the program can be retrieved from the webpage \url{http://home.iaa.es/~epm/HII-CHI-mistry.html}} framework was originally developed as a Bayesian-like methodology based on the comparison between observed emission-line ratios and extensive grids of photoionisation models.
Recent developments of the code extended it   towards sulphur determinations in star-forming galaxies \citep{pm25b}, demonstrating that optical
sulphur diagnostics can provide robust abundance constraints over wide
metallicity ranges while reproducing the expected abundance trends
derived from large spectroscopic samples,
in good agreement with direct-method abundances, which constitutes one of the main strengths of the methodology.

The use of photoionisation models to derive total S abundances is especially convenient as the main observational difficulty  using optical lines arises from the incomplete ionic
coverage provided by optical spectroscopy. The strongest optical and near-IR sulphur
lines, [\sii] $\lambda\lambda$6717,6731 and [\siii] $\lambda\lambda$9069,9532, trace S$^+$ and S$^{2+}$, but higher ionisation stages such as S$^{3+}$, expected to become increasingly
important under hard radiation fields, emit primarily in the mid-IR
through transitions such as [\siv] 10.5 $\mu$m \citep{vermeij02}. 
Consequently, total sulphur abundance determinations based only on optical spectroscopy
traditionally rely on ionisation correction factors (ICFs) \citep{stasinska78,pm06,dors16}.
Classical sulphur ICF prescriptions were primarily calibrated under
\hii\ region conditions, where oxygen and sulphur ionisation structures remain relatively coupled \citep{pc69}. However, AGN photoionisation conditions differ substantially. Harder ionising continua can maintain extended high-ionisation regions, potentially altering the correspondence between oxygen and sulphur ionic fractions. Under these conditions, empirical ICF prescriptions derived for stellar ionising fields may become inadequate (e.g. \citealt{monteiro21,dors23,zhu24}).

 Successive implementations of {\sc HCm} already extended the method toward different ionising sources and wavelength regimes. Particularly relevant to this work, the AGN
implementation of {\sc HCm} \citep{hcm-agn} demonstrated that model-based abundances are possible even under hard ionisation fields characteristic of active nuclei, such as double power-law continua, Post-Asymptotic Giant Branch (pAGB) evolved stars, and advection-dominated accretion flow (ADAF, \citealt{adaf}) prescriptions \citep{pd25}. 
More recent developments incorporated infrared diagnostics both for star-forming objects \citep{jafo21b} and AGN \citep{pd22}.
Infrared implementations incorporating high-ionisation transitions have highlighted the importance of properly accounting for unseen ionic stages in AGN abundance studies, including a more complete study of the total sulphur abundance \citep{pd24a}. 

These developments naturally motivate extending sulphur abundance determinations based on optical lines toward AGN environments. The present work extends {\sc HCm} towards the determination of total sulphur abundances from optical spectra in the NLR of AGN without relying on empirical sulphur ICFs. The method incorporates photoionisation grids exploring variable S/O ratios and multiple ionising spectral energy distributions (SEDs), including AGN power-law
continua, pAGB stellar ionisation and ADAF models. We test the consistency of the method against AGN with published direct-method ionic abundances and apply it to a large sample of AGN-like MaNGA spectra to investigate sulphur
abundance patterns under different ionisation conditions.

The paper is organised as follows. In Section 2 we present the sample of objects used to validate and apply our method. In Section 3 we describe the photoionisation models and the methodology used by {\sc HCm} to derive S abundances in AGN. In Section 4 we present our results and discuss them, including the analysis of the control sample to check the agreement with the results from the direct method, and the application to a large sample of MaNGA data. Finally, in Section 5 we summarise our results and present our conclusions.

\begin{figure}
   \centering
   \includegraphics[width=0.45\textwidth,clip=]{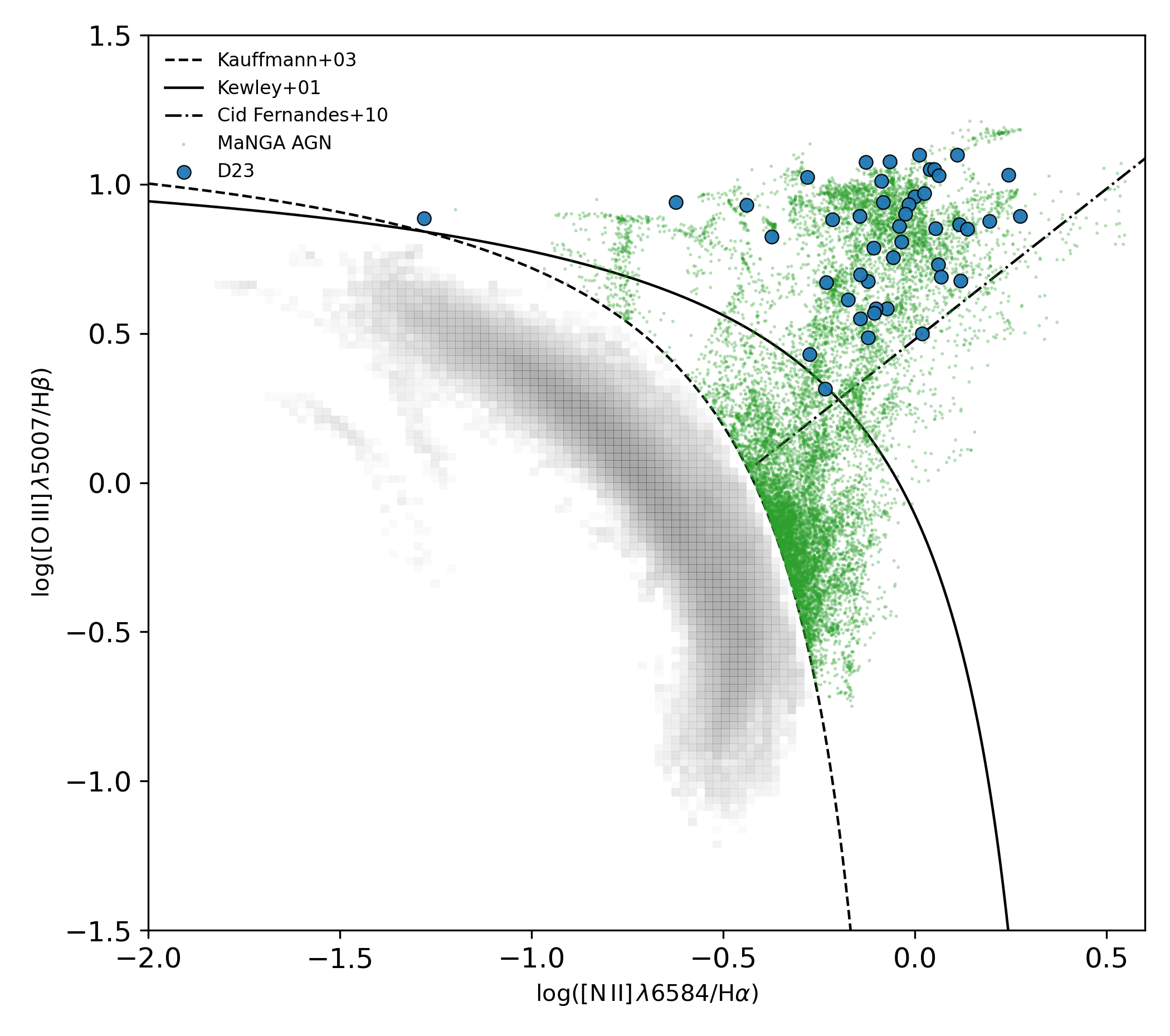}
   \caption{Classical BPT diagnostic diagram based on the emission-line ratios [\nii] $\lambda$6584/\ha\ and [\oiii]$\lambda$5007/\hb. The grey density map represents the distribution of MaNGA star-forming pointings studied in \cite{pm25b}, providing the reference locus of \hii-like ionisation. Green points correspond to the MaNGA pointings used to select the AGN analysed in this work, while blue circles indicate the control sample of nearby AGN from D23. The dashed and solid curves show the empirical demarcation of \cite{kauffman03} and the theoretical maximum starburst line of \cite{kewley01}, respectively, separating star-forming and AGN-dominated ionisation. The dot-dashed line shows the Seyfert–LINER separation of \cite{cid10}.}

   \label{bpt}
\end{figure}

\begin{figure}
   \centering
   \includegraphics[width=0.45\textwidth,clip=]{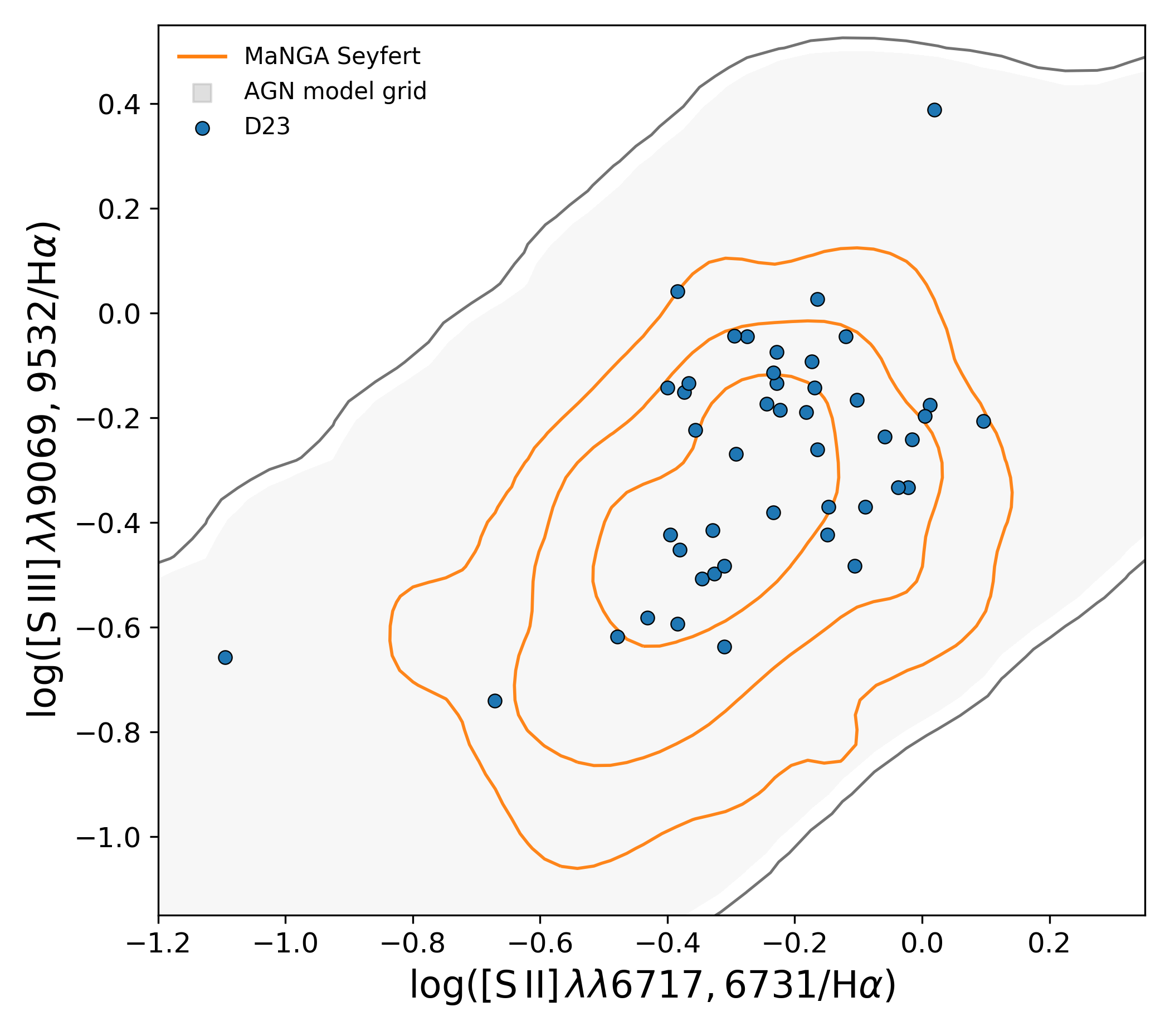}
   \caption{Sulphur diagnostic diagram comparing the MaNGA Seyfert regions (orange density contours) and the D23 control sample (blue circles). The grey shaded area represents the region of the diagram covered by the AGN photoionisation model grid adopted in this work, computed assuming an ionising continuum with $\alpha_{\rm OX}=-1.2$ and a stopping criterion of a free electron fraction of 98\%. The [\sii] emission corresponds to the sum of the $\lambda\lambda6717,6731$ doublet, while [\siii] corresponds to the sum of the $\lambda\lambda9069,9532$ doublet, both normalised to \ha.}

   \label{s2-s3}
\end{figure}

\section{Data samples}\label{sec:data}

\subsection{Control sample}\label{sec:control_sample}

As a first test of the method we used a control sample of Seyfert 2 galaxies  with reddening-corrected continuum-subtracted optical emission-line intensities and published ionic oxygen and sulphur abundances derived from electron temperature measurements by \cite{dors23} (hereafter D23).

This sample consists of 33 nearby ($z\: \la \: 0.08$) objects with public narrow (Full Width Half Maximum (FWHM) $< \: 1000\: \rm km\: s^{-1}$) optical emission lines,  through the Sloan Digital Sky Survey Data Release 17 (SDSS DR17 \citealt{sdss-dr17}) in the spectral interval 
 $3000 \: < \lambda($\AA$) \: < \: 9100$, including from [\oii] $\lambda$3727\AA\ to [\siii] $\lambda$9069\AA, and also with [\oiii]$\lambda$4363\AA\ auroral line that, along with the corresponding strong nebular [\oiii] line at $\lambda$ 5007 \AA, allows the estimation of the electron temperature and the derivation of ionic abundances following the direct method.   
The control sample contains 12 additional objects classified as Seyfert 2  from  the  literature in a more restricted spectral range, without the  presence of the [\siii]$\lambda9069$ line, which  was taken from \citet{riffel06}.

For our comparison we also took from D23 both ionic abundances for O$^+$, O$^{2+}$, S$^+$, and S$^{2+}$ as derived using the direct method from the electron temperature measured using the ratio of [\oiii] lines and the code {\sc pyneb} \citep{pyneb}. We also took total O and S elemental abundances as calculated in D23 from the assumed ICFs in that work.   

These abundances are later compared with results from our method in Section 4.1, but
the role of this control sample must be distinguished for ionic and total abundances. The ionic abundances provide an observational benchmark, since they are directly derived from the measured electron temperatures and emission-line intensities. In contrast, the total O/H and particularly S/H values reported by D23 depend on the adopted ICF prescriptions and cannot be regarded as independently known reference abundances. We therefore use the comparison with the ionic abundances as a consistency test of the {\sc HCm} solutions and of the adopted model grids, whereas the comparison of total abundances is used to investigate the differences introduced by the ionisation corrections implicit in each methodology.

\subsection{MaNGA AGN sample}\label{sec:manga_sample}

As in \cite{pm25b} with star-forming objects, a second sample of data was analysed to apply our methodology in a statistically significant sample to analyse the behaviour of the fundamental abundance relations over a wide range of AGN conditions, using strong optical [\sii] and [\siii] emission lines.
This was drawn from the Mapping Nearby Galaxies at Apache Point Observatory survey \citep[MaNGA;][]{Bundy2015}, which is part of the Sloan Digital Sky Survey IV \citep[SDSS IV;][]{Blanton2017}.

We selected from  the {\sc MaNGA} data release 17 (DR17, \citealt{sdss-dr17}) those  spaxels in the corresponding datacubes with a size significantly smaller than the point spread function (PSF), which has   a median FWHM of 2.54 arcsec (Law et al. 2016), adopted here as the effective spatial resolution. At the median redshift of our sample this corresponds to an average physical resolution of approximately 1.2 kpc.
Emission-line fluxes were collected in each selected spaxel with at least  a signal-to-noise ratio (S/N) of 10 for [\oii] $\lambda$3727 \AA, [\oiii] $\lambda\lambda$4959,5007 \AA, [\nii] $\lambda$6584 \AA, [\sii] $\lambda\lambda$6717,6731 \AA, and [\siii] $\lambda\lambda$9069,9532 \AA. Such high S/N was chosen to minimise the chance of false detections of [\siii] $\lambda\lambda$9069,9532 \AA\ lines due to possible telluric contamination of the spectrum. Moreover, to minimise the chance of false detection, we allowed fitting of  [\siii] $\lambda\lambda$9069,9532 with independent velocity dispersion of each line in doublet and then selected spectra with differences in the velocity dispersion less than 30 km/s. To obtain the emission line fluxes, we used the {\sc STARLIGHT} code \citep{CidFernandes2005,Mateus2006,Asari2007} to subtract the stellar background and the {\sc ELF3D} code to fit the emission lines. Details about the processing can be found in \citet{zinchenko2016,z21}.  
The line fluxes were corrected for interstellar reddening using the analytical approximation of the Whitford interstellar reddening law \citep{Izotov1994},
assuming the Balmer line ratio of $\text{H}\alpha/\text{H}\beta = 2.86$.
When the measured value of $\text{H}\alpha/\text{H}\beta$ is less than 2.86,
the reddening is set to zero.

Our final AGN sample contains 6\,108 spectra adopting the [\nii]/\ha\ vs. [\oiii]/\hb\ diagram from \cite{bpt} and the theoretical curve between star-forming galaxies and AGN postulated by \cite{kewley01}, as shown in Figure \ref{bpt}. 
This choice also minimises the potential impact of diffuse ionized gas (DIG), which can enhance the low-ionization [\sii] emission in star-forming regions and therefore affect their location in the [S II]-based BPT diagram. As a consistency check, we also classified the regions using the [\sii]-based diagnostic and found that 91.6\% of the sources classified as AGN from the [N II] diagram remain in the AGN region of the [\sii] diagram. We therefore conclude that our AGN selection is robust against the choice of BPT diagnostic.
We ruled out those objects classified as {\em composite} galaxies as defined by \cite{kewley06}, above the empirical demarcation curve defined by \cite{kauffman03}, in order to ensure that our sample only contains pure-AGN photoionisation. 
The sample was later divided into Seyfert 2 (5\,246 objects) and LINER-like (862 objects) classes using the separation line proposed by \cite{cid10}. This separation is essential because the two populations may be associated with different ionising sources, ionisation parameters, gas densities, and spatial scales.
The selected pointings were distributed among 218 unique galaxies. Of these, 178 galaxies host Seyfert 2 regions and 69 host LINER regions, with 29 galaxies contributing to both subsamples. The galaxies span approximately \(0 < z < 0.082\), with a median redshift of \(z \simeq 0.034\).

We also show in figure \ref{bpt} a comparison between the location of both the control sample and the selected MaNGA regions in the classical BPT diagnostic diagram. 
As can be seen, the D23 galaxies occupy the Seyfert locus, although they do not exactly reproduce the distribution of the MaNGA Seyfert sample. In particular, the D23 objects preferentially populate the high-excitation end of the AGN sequence, whereas the MaNGA sample extends towards lower ionisation conditions. However, as shown in Figure \ref{s2-s3}, when considering sulphur-based observables, both samples display a remarkable degree of overlap. In the $\log$([S II]/H$\alpha$) versus $\log$([S III]/H$\alpha$) plane, the D23 galaxies follow the same sequence defined by the MaNGA Seyferts and span a very similar range of sulphur line ratios. This agreement supports the use of the D23 sample as an external benchmark for validating the sulphur abundance determinations, while the MaNGA Seyfert sample extends the analysis over a much broader region of the observed parameter space.

\section{The models and the code}\label{sec:models}

\subsection{Photoionisation model grids}\label{sec:photoionisation_grids}

Our method to calculate abundances in the NLR of AGN is fed with a large grid of photoionisation models calculated using the code {\sc Cloudy} v. 17.02 \citep{cloudy} under different input conditions, whose predicted emission-line fluxes are later compared with an observational input. 

A first set of models,  described in \cite{hcm-agn} and later complemented in \cite{pd25}, 
was computed with varying input O/H, N/O, and log $U_*$\footnote{Notice that for the derivation of the ionisation parameter $U_*$ from {\sc HCm} we use the notation $U_*$ to underline that this parameter was derived at a fixed assumed SED and geometry, which are factors that also contribute to the variation of emission-line ratios between consecutive ionisation stages. For more details see \cite{pm25b}.} in order to let the code estimate these quantities. For the case of AGNs, the default chemical abundance range covers $12+\log({\rm O/H})$ from 6.9 to 9.1, with a step of 0.1 dex. All the rest of chemical species are scaled to the solar proportions given by \cite{asplund}, with the exception of N, which is considered in fractions of log(N/O) from -2.0 to 0.0 in steps of 0.25 dex.
The ionisation parameter spans $\log U_*$ from $-4.0$ to $-0.5$. 

In addition, as in the case of \cite{pm25b} for star-forming galaxies with the aim of providing solutions for the total sulphur abundance, we computed for AGN additional grids of  models considering several values of S/O around the solar value, allowing the code to explore S/O variations from -0.6 up to 0.6 dex in steps of 0.2 dex around the assumed solar ratio (S/O)$_{\odot}$ = -1.57 \citep{asplund}. Given that S/H determination does not depend on N lines, the models of this second grid do not present any variation in N/O as they consider the expected O/H - N/O relation described in \cite{hcm14}.   
The considered AGN models assume different families of incident SED, including double-peaked power-law with index $\alpha_{UV}$ = -1.5, and $\alpha_{ox}$ from -2.0 to -0.8 in bins of 0.2. In addition, we computed models considering post-AGB stars using SEDs from \cite{rauch} with different effective temperatures, $T_:*$ of 50, 100 and 150 kK. Finally, we also incorporated models using an Advection Dominated Accretion Flow (ADAF) from \cite{adaf}.
All models assume a filling factor of 0.1 and a standard Milky-Way gas-to-dust mass ratio, at two different stopping criteria, for 98\% and 2\% fraction of free electrons at the outermost radius.
Electron density was assumed in all models to be constant at a value 500 cm$^{-3}$. As explored in \cite{hcm-agn}, adopting values of up to 2\,000 cm$^{-3}$ does not introduce differences in the final obtained chemical abundances larger than the obtained errors.

\subsection{Description of the code {\sc HCm}}\label{sec:code_philosophy}

The new implementation of {\sc HCm} described here (v. 6.2) for optical spectra to derive total sulphur abundances in AGN follows the same general philosophy as previous versions. The code compares a set of input observed reddening-corrected emission-line ratios and their corresponding errors with the predictions made by different large grids of photoionisation models with varying conditions. For each model, a $\chi^2$-like quantity is computed from several available observables based on line ratios dependent on certain gas-phase chemical   abundances, such as O/H or N/O, and on the ionisation parameter,  $U_*$. 
The final estimates are obtained as weighted means over the model grid, using the inverse of the \(\chi^2\)-like statistic defined in \cite{hcm14} to weight each model according to its agreement with the observed emission-line ratios.

The code can use as observational input different nebular collisionally-excited lines available in the optical spectrum relative to \hb\,  including [\oii] $\lambda3727$, [\neiii] $\lambda3868$,[\oiii] $\lambda\lambda4959,5007$, [\nii] $\lambda6584$, [\sii] $\lambda\lambda6717,6731$, and [\siii] $\lambda\lambda9069,9532$. 
When auroral lines are available, such as [\oiii] $\lambda4363$ or [\nii] $\lambda5755$, their ratios with the corresponding nebular lines provide additional constraints, but they are not mandatory for the calculation.

Departing from the complete grid, the code performs several consecutive iterations through more constrained grids of models to derive abundances, given the simultaneous dependence of some of the chosen observables on different elemental abundances.  
The first iteration, for the determination of N/O,  and the second iteration, for O/H and $U_*$, are well described in the case of AGN in \cite{hcm-agn} and \cite{pd25}.

A third iteration, already implemented in v. 6.1 for the case of star-forming objects, was implemented to calculate S/H using grids of models with variable S/O ratio, conveniently constrained once O/H and $U_*$ have been estimated in the previous iteration. However, before that and as it is described in \cite{pm25b} for  objects ionised by massive stars, the inclusion of [\siii] among the analysed emission lines defines new observables that can better be used in the second iteration to derive O/H and $U_*$ also for the case of AGN.

\begin{figure*}
   \centering
   \includegraphics[width=0.45\textwidth,clip=]{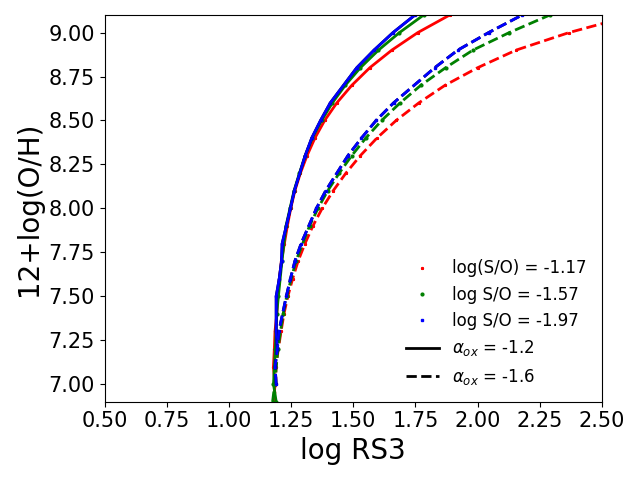}
   \includegraphics[width=0.45\textwidth,clip=]{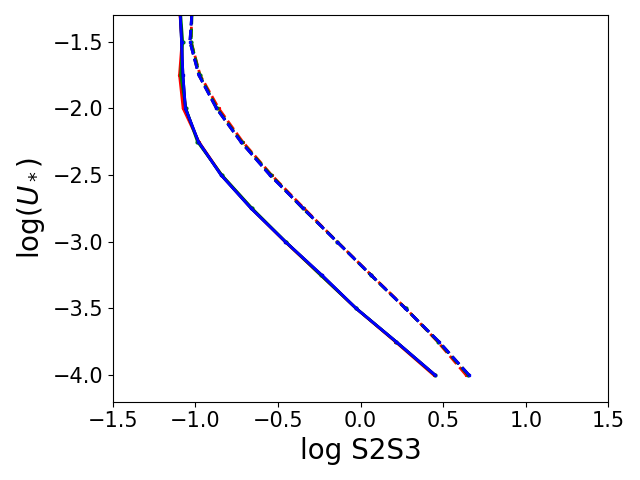}
   \caption{Predicted relation between certain observables based on [\siii] emission lines and derived properties for different assumed S/O ratios. Left: Relation between $RS3$ and total oxygen metallicity for fixed log $U_*$ = -2.0. Right: Relation between $S2S3$ and the ionisation parameter. In both panels lines join models calculated assuming an incident SED with a double-peaked power-law with index $\alpha_{\textrm ox}$ = -1.2 (solid) and  -1.6 (dashed).}

   \label{RS3}
\end{figure*}

\begin{figure*}
   \centering
   \includegraphics[width=0.45\textwidth,clip=]{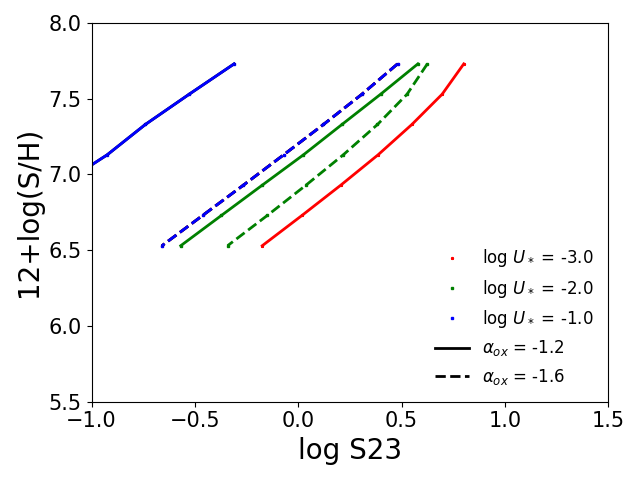}
   \includegraphics[width=0.45\textwidth,clip=]{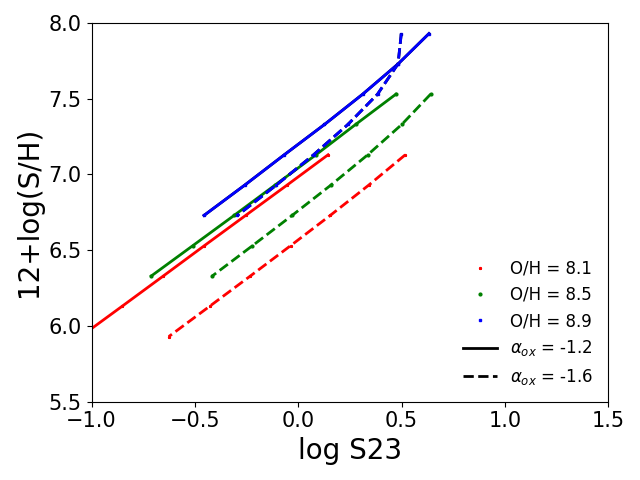}
   \caption{Relation between the S23 parameter and the total sulphur abundance as predicted by several sequences of photoionisation models assuming a double-peaked power-law for AGN with $\alpha_{\textrm ox}$ = -1.2 (solid line) and -1.6 (dashed line) calculated assuming a stopping criterion of $f_e$ = 98\%. The left and right panels show variations in log $U_*$ and input O/H, respectively.} 
   \label{S23}
\end{figure*}

Among these observables based on [\siii] lines is the $RS3$ parameter, defined as:

\begin{equation}
RS3 = \frac{\mathrm{[S\,III]} 9069,9532 \AA}{\mathrm{[S\,III]} 6312 \AA}
\end{equation}    

In order to illustrate the dependence of $RS3$ on the oxygen abundance, and to motivate its use as a constraint in the fitting procedure,  left panel of Fig. \ref{RS3} shows the relation between O/H and $RS3$, where we show sequences of models for two different values of the assumed $\alpha_{\textrm{ox}}$ = -1.2 and -1.6, at a fixed log $U_*$ value of -2.0, and for different values of S/O around the solar ratio.
 As expected, despite its dependence on the assumed SED, $RS3$ shows a monotonic dependence on O/H over the full range covered by the models, confirming its suitability as a metallicity tracer in AGN environments. In contrast, its dependence on S/O is significantly weaker. For a variation of almost 0.8 dex in log(S/O), the resulting displacement in $RS3$ remains relatively small compared to the overall variation produced by changes in O/H. This behaviour indicates that $RS3$ is primarily driven by total gas-phase metallicity, while the influence of the sulphur-to-oxygen ratio acts only as a second-order effect. At low metallicities ($12+\log({\rm O/H}) \lesssim 7.6$), the model sequences become progressively steeper, implying a reduced sensitivity of $RS3$ to abundance variations,  a behaviour similar to that previously found for the $RO3$ (i.e. the corresponding ratio for [\oiii] lines) parameter in {\sc HCm} for AGN  \citep{hcm-agn}. The effect of the ionising SED is also moderate. Models with a softer continuum ($\alpha_{\rm ox}=-1.6$) are systematically shifted towards slightly larger RS3 values at a given oxygen abundance with respect to the $\alpha_{\rm ox}=-1.2$ models, although the displacement remains considerably smaller than the variation produced by O/H across the grid. Therefore, while RS3 retains some dependence on both S/O and the adopted ionising continuum, its dominant sensitivity is to the overall metallicity of the gas.

Another observable based on [\siii] lines that is better utilised in the second iteration of the code aimed at the determination of both O/H and $U_*$ is the $S2S3$ parameter, defined as:

\begin{equation}
S2S3  =  \frac{\mathrm{[S\,II]} 6717,6731 \AA}{\mathrm{[S\,III]} 9069,9532 \AA}
\end{equation}

This parameter is often used as a proxy for the ionisation parameter \citep{diaz98} for star-forming objects, but its validity for AGN must also be explored.
The right panel of Figure \ref{RS3} shows the relation of this parameter with log $U_*$ for the same model sequences at a fixed 12+log(O/H) = 8.7 and different assumed S/O values and for two different values of the SED shape.
 In contrast to RS3, S2S3 is primarily controlled by the ionisation state of the gas. Variations in S/O produce only minor displacements of the model sequences, while the effect of the adopted ionising continuum is similarly small. The strongest dependence is therefore on the ionisation parameter itself, confirming that S2S3 provides an efficient tracer of the ionisation conditions in AGN narrow-line regions. This behaviour is expected because the parameter compares two consecutive ionisation stages of the same element and is therefore largely insensitive to the overall chemical abundance.

The behaviour of S2S3 is complementary to that of the O2O3 parameter discussed by \citet{hcm-agn}. Although both parameters are strongly sensitive to the ionisation parameter, their dependence on $\log(U)$ changes at different ionisation regimes. In the case of O2O3, the turnover occurs around $\log(U)\simeq-2.5$, whereas for S2S3 it takes place closer to $\log(U)\simeq-2$. As a consequence, the regions where each parameter becomes less sensitive to changes in the ionisation parameter do not coincide. The simultaneous use of O2O3 and S2S3 therefore provides a more robust determination of $\log(U)$ than either parameter alone, reducing degeneracies and improving the constraints on the ionisation conditions adopted by {\sc HCm}.

Once the code computes O/H and $U_*$ in this second iteration, these can be used as constraints for a new iteration of the code to independently derive S/H over the grid of models with variable S/O. In this case, the only observable used by the code to compare observed emission-lines and models is $S23$, defined by \cite{ve96} as:

\begin{equation}
S23 = \frac{\mathrm{[S\,II]} 6717,6731 \AA + \mathrm{[S\,III]} 9069,9532 \AA}{H\beta}
\end{equation}

although, in the absence of the nebular [\siii] lines, the code can also define an observable based on [\siii] $\lambda$ 6312 \AA\ auroral line. This parameter has been used to empirically calibrate both total O/H \citep{dpm00, pmd05} and S/H \citep{pm06}.
Indeed, it presents a non-negligible dependence on both abundances, as already demonstrated for star-forming galaxies by
\cite{pm25b}. This is also the case for AGN, as shown in Fig. \ref{S23}, where it can be seen that this parameter depends, in addition to the SED shape, simultaneously on the total metallicity, given the collisional nature of the involved emission lines, and on sulphur abundance. Therefore, in order to derive S/H from $S23$, a previous estimation of O/H and U is advisable prior to any trial to derive S abundance using these lines. 

This necessary step for the optical version of the code contrasts with the strategy designed for the code in the IR in \cite{pd24a} where S lines can be used to estimate S/H abundances without any previous determination of the total metallicity, as IR lines have very little dependence on electron temperature. On the other hand, for the optical version described here, if [\sii] and [\siii] lines are given to the code, this will provide both a determination for O/H, $U_*$, and S/H.

As a quick test of the validity of the code to retrieve elemental abundances using only as input the strongest emission-line fluxes, we checked that using as input the relative emission lines of [\oii] $\lambda$3727\AA, [\oiii]$\lambda$5007 \AA, [SII]$\lambda$6725 \AA\ and [\siii]$\lambda$ 9069 \AA, with an additional uncertainty of 10\%, we obtain O/H with a mean offset of 0.05 dex and S/H with 0.08 dex for all the assumed SEDs and geometries. These offsets are even smaller when all lines, including auroral lines, are included, although the derived abundances may deviate when an inappropriate SED is adopted, introducing an additional source of uncertainty.

\section{Results and discussion}

\subsection{Validation with the control sample}\label{sec:control}

\begin{table*}
\begin{center}
\caption{Mean offsets ($\Delta$) and standard deviations ($\sigma$) of the comparison between the ionic abundances as derived using the direct method by D23 and the estimations made by {\sc HCm} as described in the text assuming different $\alpha_{\textrm{ox}}$ for the incident SED and different stopping criteria for the models. The offsets for the comparison for total abundances are also listed.} 

\begin{tabular}{lccccccc}
\hline
$\alpha_{\textrm{ox}}$ & $f_e$ & $\Delta$(O$^+$+O$^{2+}$) & $\sigma$(O$^+$+O$^{2+}$) & $\Delta$(S$^+$+S$^{2+}$) & $\sigma$(S$^+$+S$^{2+}$) & $\Delta$(O/H) & $\Delta$(S/H) \\
\hline
-0.8   &   0.02 & -0.72  &  0.29  & +0.07  &  0.18  & -0.18  &  -0.12 \\
-0.8   &   0.98 & +0.08  &  0.20  & -0.12  &  0.19  & +0.16  &  +0.18  \\
-1.0   &   0.02 & -0.15  &  0.17  & +0.19  &  0.17  & +0.22  &  +0.18 \\
-1.0   &   0.98 & +0.09  &  0.18  & -0.10  &  0.22  & +0.19  &  +0.26  \\
-1.2   &   0.02 & -0.23  &  0.17  & +0.20  &  0.16  & +0.15  &  +0.20 \\
{\bf -1.2}   &   {\bf 0.98} & {\bf -0.03}  &  {\bf 0.20}  & {\bf +0.03}  &  {\bf 0.27}  & {\bf +0.03}  &  {\bf +0.31}  \\
-1.4   &   0.02 & +0.12  &  0.16  & +0.39  &  0.29  & +0.24  &  +0.39 \\
-1.4   &   0.98 & +0.28  &  0.19  & +0.39  &  0.27  & +0.20  &  +0.48  \\
-1.6   &   0.02 & +0.29  &  0.16  & +0.54  &  0.25  & +0.24  &  +0.44 \\
-1.6   &   0.98 & +0.34  &  0.20  & +0.62  &  0.30  & +0.20  &  +0.52  \\
-1.8   &   0.02 & +0.38  &  0.17  & +0.67  &  0.28  & +0.23  &  +0.54 \\
-1.8   &   0.98 & +0.36  &  0.20  & +0.65  &  0.32  & +0.20  &  +0.47  \\
-2.0   &   0.02 & +0.39  &  0.18  & +0.63  &  0.33  & +0.23  &  +0.46 \\
-2.0   &   0.98 & +0.37  &  0.20  & +0.66  &  0.33  & +0.20  &  +0.45  \\
\hline

\label{table_d23}
\end{tabular}
\end{center}
\end{table*}

\begin{figure*}
   \centering
   \includegraphics[width=0.45\textwidth,clip=]{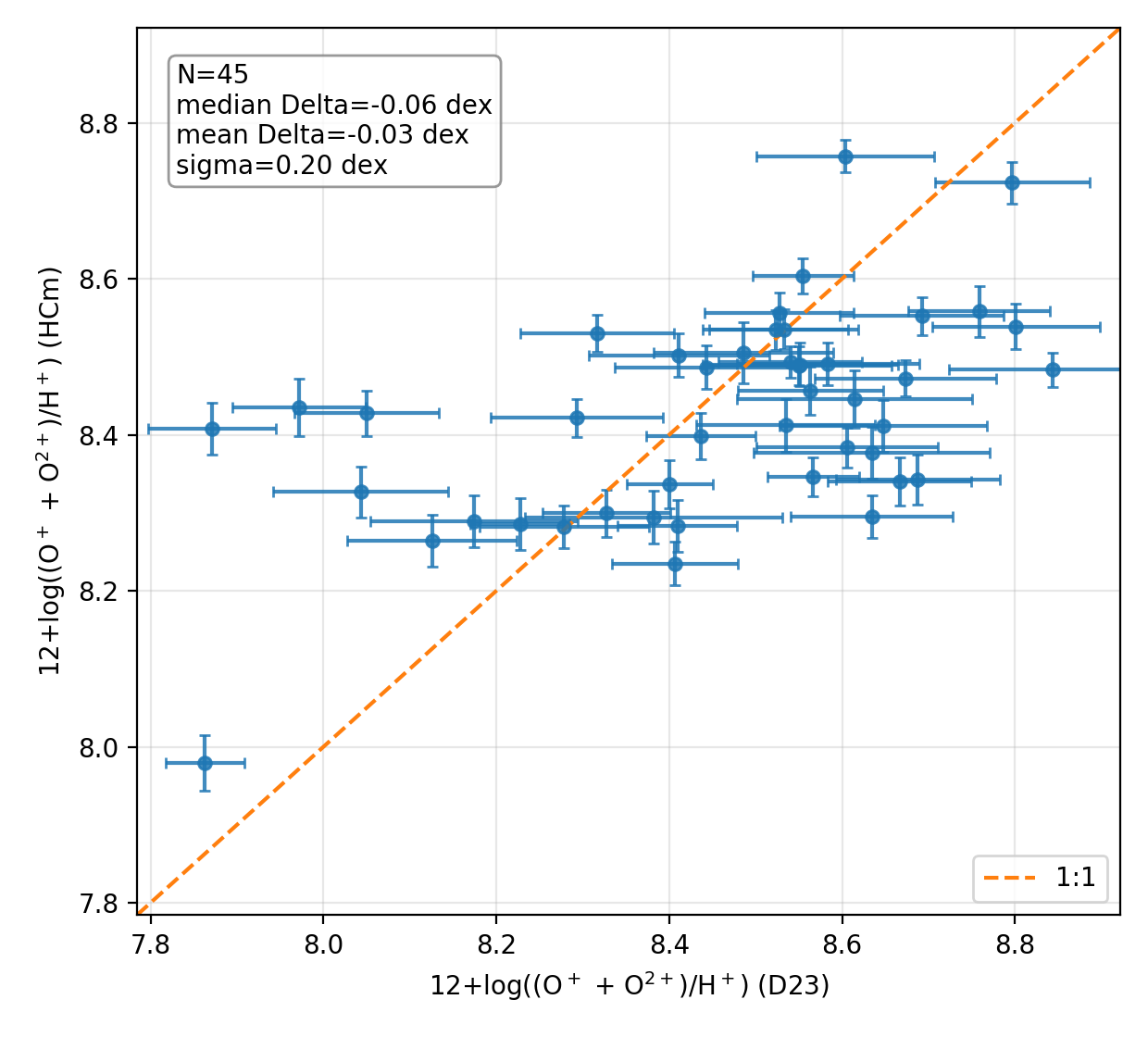}
   \includegraphics[width=0.45\textwidth,clip=]{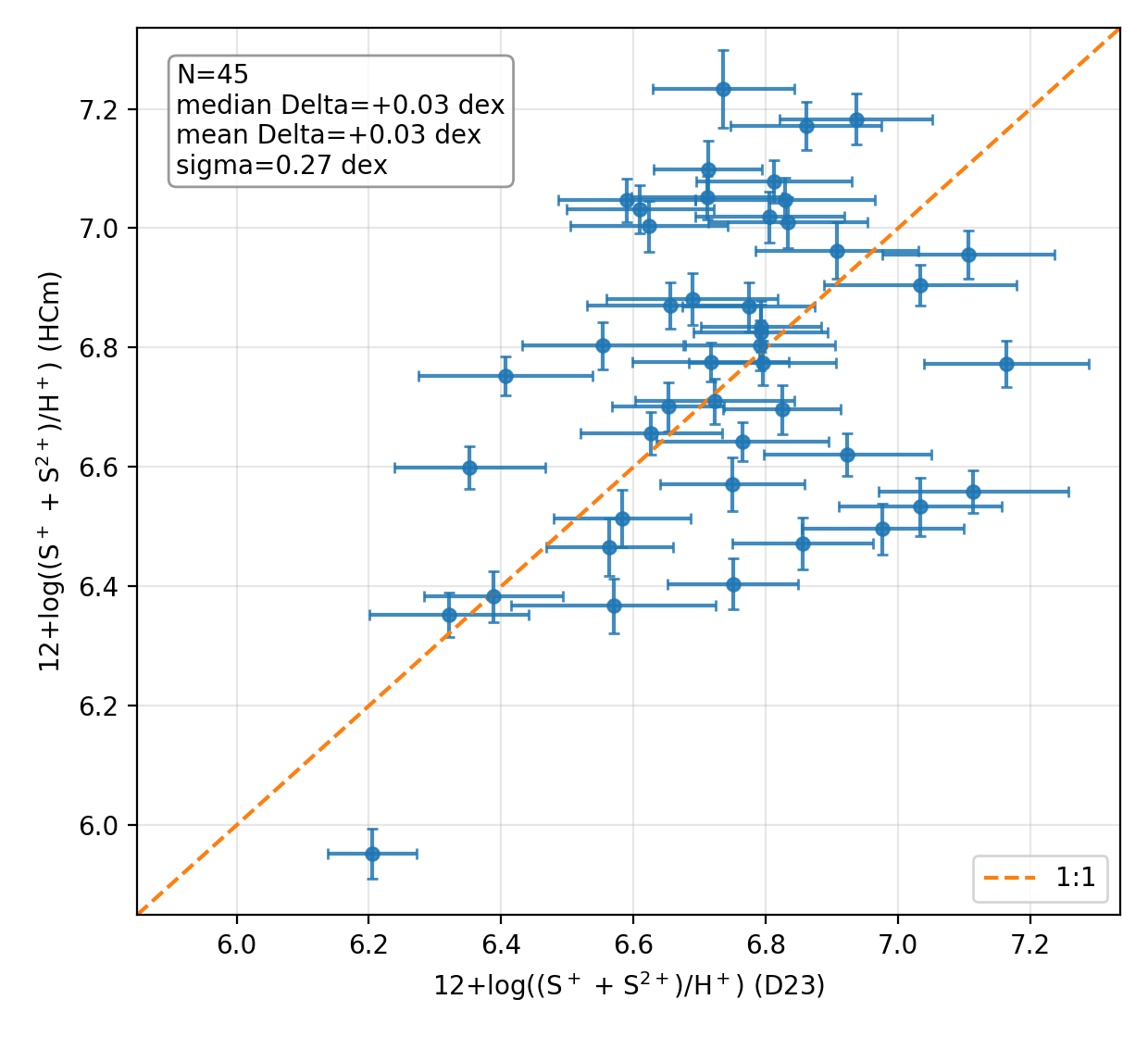}
   \caption{Comparison between the observed ionic abundances derived from the direct sum of ionic species in D23 with the ionic-equivalent abundances inferred from the {\sc HCm} total-abundance solutions and the photoionisation model grid for $\alpha_{ox}$ = -1.2 and a stopping criterion corresponding to a free electron fraction of 98\%. Left panel: oxygen, comparing the observed $12+\log((\mathrm{O}^{+}+\mathrm{O}^{2+})/\mathrm{H}^{+})$ values with the ionic-equivalent abundances reconstructed from the {\sc HCm} O/H and ionisation parameter solutions. Right panel: same comparison for sulphur using $12+\log((\mathrm{S}^{+}+\mathrm{S}^{2+})/\mathrm{H}^{+})$. The dashed line represents the one-to-one relation. Error bars on the horizontal axis correspond to the propagated observational uncertainties of the ionic sums, while vertical error bars represent the dispersion of compatible models within the adopted three-dimensional parameter space in O/H, S/H, and $\log U_*$. The inset boxes indicate the number of objects, median offset, mean offset, and standard deviation of the residuals between the model-grid ionic-equivalent abundances and the observed ionic abundances.}

   \label{comp_ionic}
\end{figure*}

\begin{figure*}
   \centering
   \includegraphics[width=0.45\textwidth,clip=]{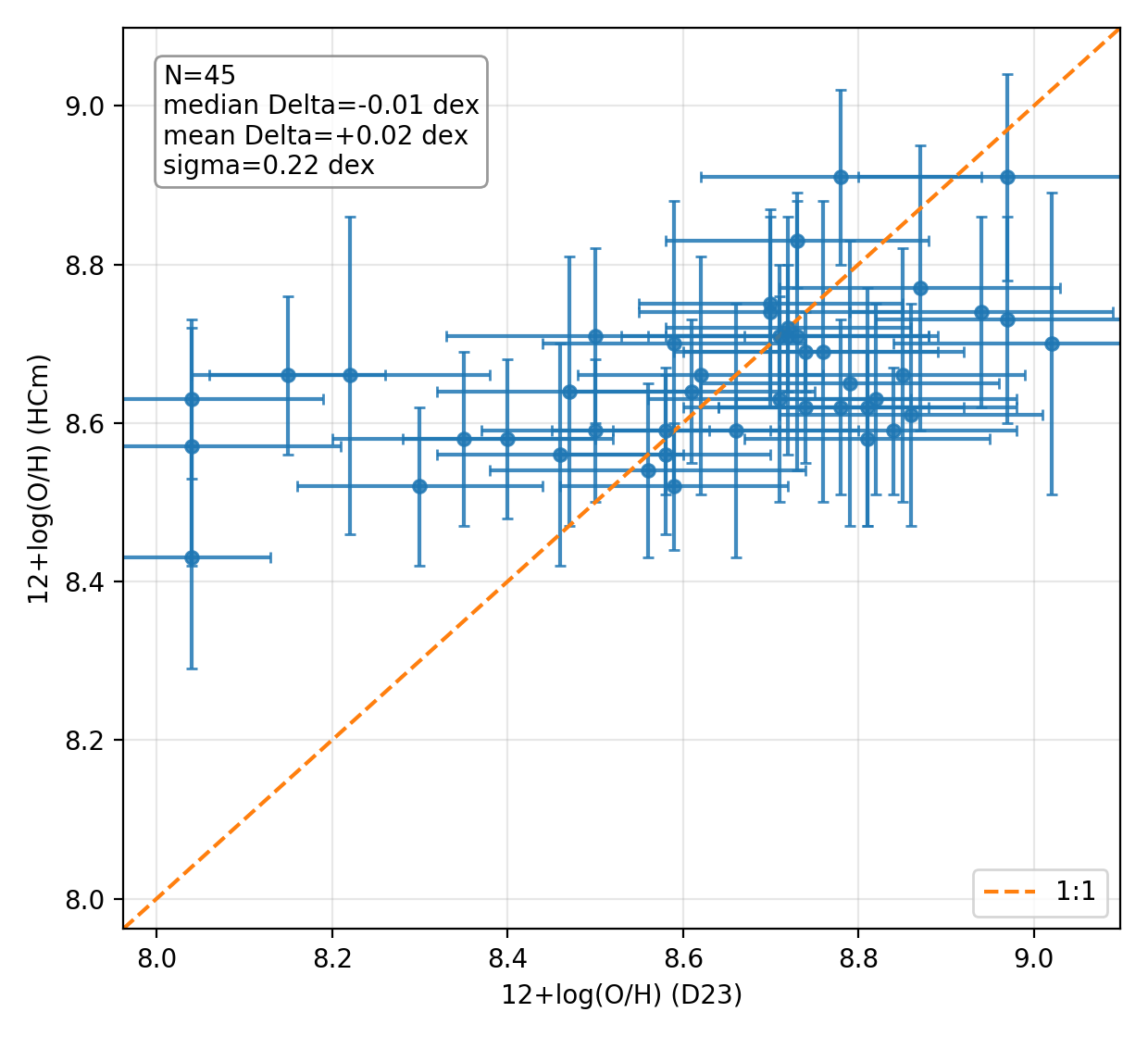}
   \includegraphics[width=0.45\textwidth,clip=]{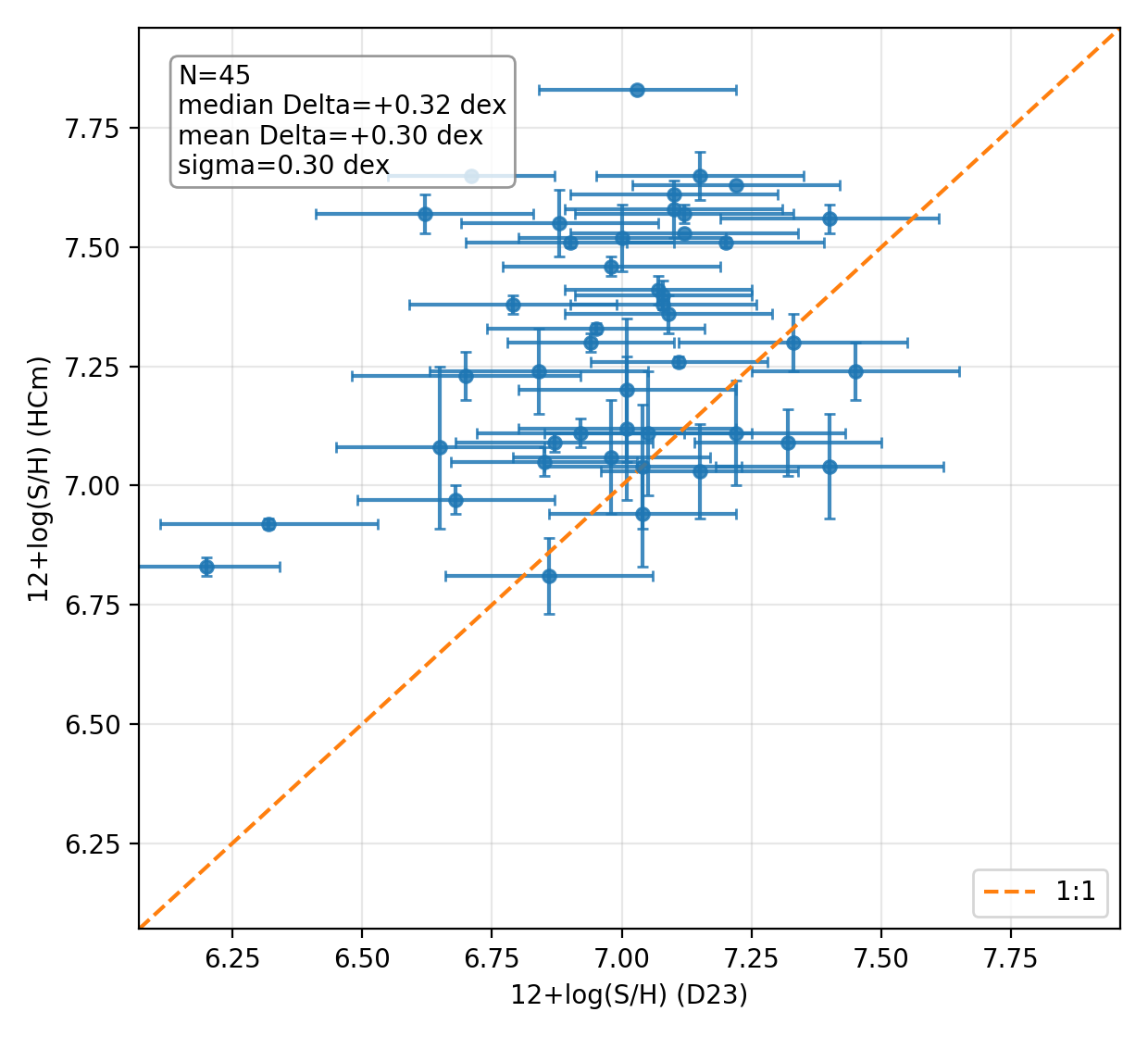}
   \caption{Comparison between total oxygen (left) and sulphur (right) abundances as reported by D23 and as obtained from {\sc HCm} assuming an incident SED with a double-peaked power-law with $\alpha_{\textrm{ox}}$ = -1.2 and a stopping criterion with a fraction of free electrons of 98\%. Symbols are the same as in Fig. \ref{comp_ionic}.}

   \label{comp_tot}
\end{figure*}

The full D23 sample of 45 objects provides a consistency test of the {\sc HCm} solutions under AGN ionisation conditions.
In particular, the electron temperatures measured by D23 allow ionic abundances to be derived independently of the ICF prescriptions required to estimate total abundances. For each object, {\sc HCm} was run using the reddening-corrected optical emission lines and model grids spanning different values of $\alpha_{\rm ox}$ and different stopping criteria. We then used the photoionisation models contributing to each {\sc HCm} solution to calculate the ionic fractions associated with the inferred physical conditions. This allows us to compare the ionic abundances implied by the {\sc HCm} model solutions with those independently derived from the measured electron temperatures, without using the D23 total abundances as constraints.

Table \ref{table_d23} summarises the mean offsets and dispersions obtained for the different model assumptions. 
Among the explored grids, the best overall agreement is obtained for \(\alpha_{\rm ox}=-1.2\) and a stopping criterion corresponding to a free-electron fraction of 98\%. In this configuration, the mean residuals are only -0.03 dex for oxygen and +0.03 dex for sulphur ionic abundances, compared with (+0.09, -0.10) dex for \(\alpha_{\rm ox}=-1.0\) and (+0.28, +0.39) dex for \(\alpha_{\rm ox}=-1.4\). A paired bootstrap resampling of the 45 objects in the control sample, using the joint distance of the mean oxygen and sulphur residuals from zero, identifies \(\alpha_{\rm ox}=-1.2\) as the least biased of these three grids in approximately 94\% of the realisations.
In any case, these values should be regarded as representative of the control sample rather than as a universal optimum for individual AGN, since the shape of the ionising continuum may vary from object to object depending on the physical conditions of the accretion disc and corona.

Figure \ref{comp_ionic} illustrates this comparison for the preferred model grid. The overall agreement is satisfactory  considering the complexity of AGN ionisation structures and confirms that the {\sc HCm} methodology remains capable of reproducing direct-method ionic abundances even under hard ionising radiation fields. 

The behaviour summarised in Table \ref{table_d23} also provides additional physical information. Moving away from $\alpha_{\rm ox}=-1.2$ produces systematic trends in both oxygen and sulphur residuals. Softer continua progressively overestimate the ionic sulphur abundance, whereas harder continua tend to underestimate it. The simultaneous minimisation of the oxygen and sulphur residuals therefore independently favours an ionising continuum close to $\alpha_{\rm ox}=-1.2$, providing empirical support for the choice adopted throughout the rest of this work.

In addition, we show in Figure \ref{s2-s3} a comparison between the observational sulphur diagnostic diagram with a shaded area that represents the region covered by the photoionisation models adopting the $\alpha_{\textrm ox}$ and geometry  that provides the closest agreement with the ionic oxygen and sulphur abundances derived for the D23 calibration sample. 
 The grid spans nearly the entire region occupied by both the MaNGA Seyfert sample and the D23 galaxies, encompassing the full observed range of [S II]/H$\alpha$ ratios and the vast majority of the measured [\siii]/\ha\ values. Only a few isolated D23 objects lie close to or slightly beyond the model boundaries. The broad overlap between the observational and model parameter spaces demonstrates that the adopted grid provides an appropriate basis for deriving sulphur abundances over the range of AGN conditions explored in this work.

Having established the consistency of the {\sc HCm} solutions with the independently derived ionic abundances, we can separately compare the total abundances inferred by {\sc HCm} with those reported by D23. Unlike the ionic-abundance comparison, this does not constitute a comparison against independently known total abundances, since the D23 values themselves rely on ICF prescriptions. 
Figure \ref{comp_tot} shows that the total oxygen abundances derived by {\sc HCm} display only a small mean offset relative to those reported by D23, but the residuals are not uniformly distributed over the full abundance range. In particular, for objects with 12+log(O/H)$_{\rm D23}\lesssim 8.5$, {\sc HCm} systematically yields higher abundances, clustered around a relatively narrow, slightly subsolar range. At higher O/H the agreement is closer, although considerable object-to-object scatter remains. This behaviour indicates that the small global mean offset should not be interpreted as evidence for a one-to-one agreement between the two total oxygen abundance scales.

Sulphur abundances, however, display a much larger systematic offset. The {\sc HCm} solutions exceed the direct-method plus ICF determinations by approximately 0.3 dex on average.
This discrepancy strongly suggests that the classical sulphur ionisation correction factors commonly adopted for \hii\ regions underestimate the contribution of higher sulphur ionisation stages under AGN conditions. While the direct method accurately measures S$^{+}$ and S$^{2+}$, 
the correction for unobserved higher ionisation stages of sulphur becomes increasingly uncertain in the presence of hard ionising continua. In AGN, the sulphur ionisation structure is not expected to follow the same relations observed in stellar photoionised nebulae, making empirical ICF prescriptions calibrated on \hii\ regions potentially inadequate.

This comparison indicates that the discrepancy is not driven by the ionic abundances themselves, 			but by the ionisation correction applied to recover the total sulphur abundance. In particular, the classical assumption that the ionisation structure of sulphur can be directly traced by that of oxygen, as adopted in direct-method analyses, does not appear to hold for AGN-like ionising spectra.
The failure of this assumption is expected in AGN environments, where the harder ionising continuum can produce a significant fraction of sulphur in ionisation stages higher than S$^{2+}$, without a one-to-one correspondence with the O$^+$/O$^{2+}$ balance.

The control sample therefore provides two main results. First, the ionic abundances implied by the preferred {\sc HCm} model grid are consistent, on average, with those independently derived from electron-temperature measurements, although this agreement does not imply that the corresponding total abundances are uniquely determined. In particular, the comparison of total O/H reveals a systematic compression towards slightly subsolar {\sc HCm} values at the low-O/H end of the D23 sample. Second, the substantially larger and systematic excess in total S/H obtained with {\sc HCm} relative to the ICF-based D23 values points to important differences in the treatment of the unobserved sulphur ionisation stages. Rather than identifying either total-abundance scale as the true one, this comparison indicates that classical ICF prescriptions calibrated under stellar ionisation conditions may underestimate S/H when applied to AGN.

The adopted sampling of \(\Delta\alpha_{\rm ox}=0.2\) does not allow the optimum SED to be constrained more precisely than this interval, and an optimum value between adjacent grid points cannot be excluded. However, given the dispersion of the individual residuals, interpolation between the neighbouring grids would not provide a statistically meaningful improvement. We therefore adopt \(\alpha_{\rm ox}=-1.2\) as the representative grid providing the best simultaneous agreement for the oxygen and sulphur ionic abundances, rather than as a precise determination of the intrinsic \(\alpha_{\rm ox}\) of the sources. Importantly, the systematic excess in total S/H relative to the D23 determinations is preserved for the neighbouring grids, with mean offsets of +0.26, +0.31, and +0.48 dex for \(\alpha_{\rm ox}=-1.0,-1.2,\) and -1.4, respectively.

\begin{figure}
   \centering
   \includegraphics[width=0.45\textwidth,clip=]{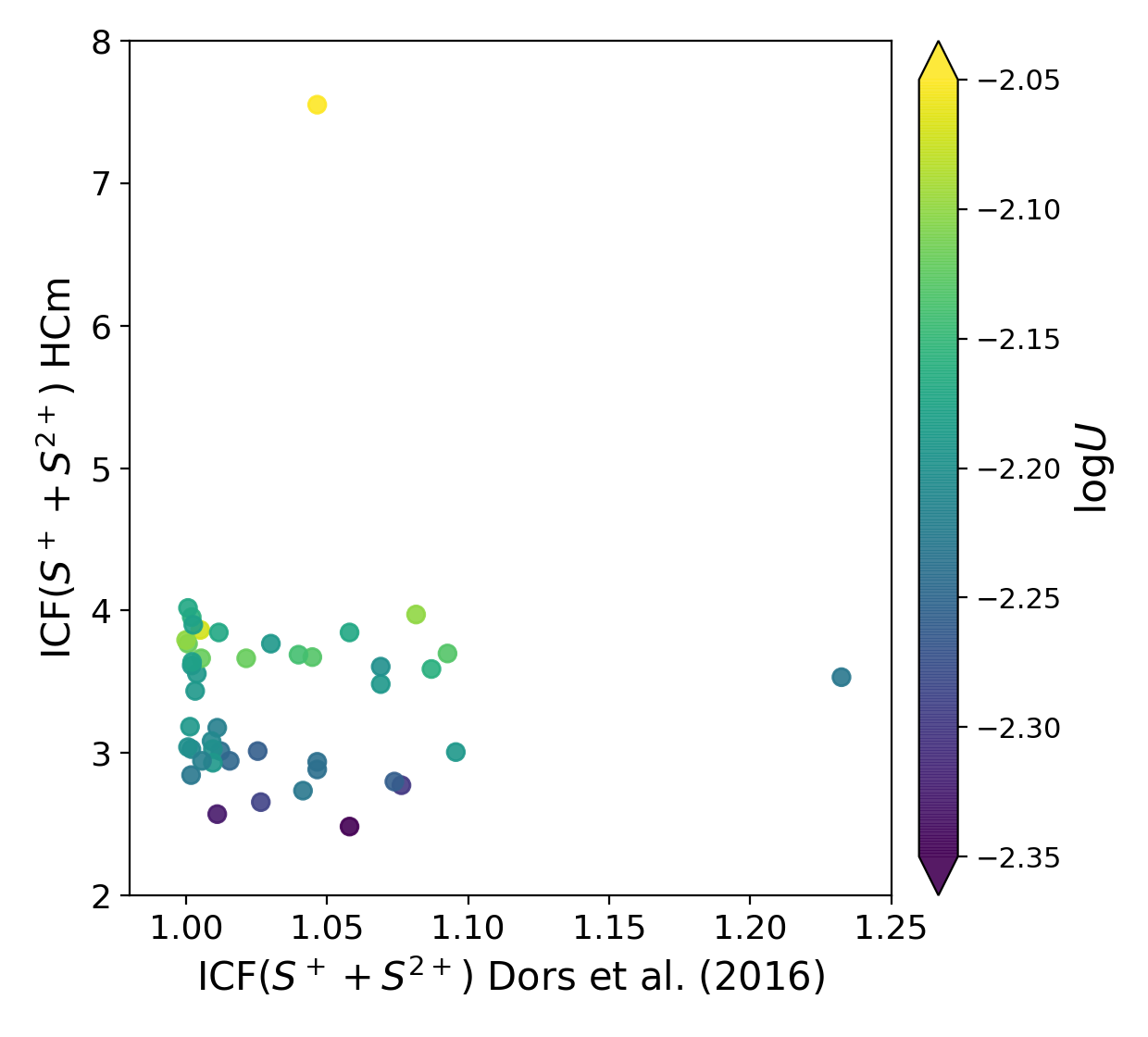}
\caption{Comparison between the sulphur ionisation correction factor, $\mathrm{ICF}(\mathrm{S}^{+}+\mathrm{S}^{2+})$, defined as $(\mathrm{S/H})/[(\mathrm{S}^{+}+\mathrm{S}^{2+})/\mathrm{H}^{+}]$, predicted by the AGN photoionisation models used by {\sc HCm} and that derived from the prescription of \cite{dors16} for the objects in the control sample.
The latter is calculated from the observed $\mathrm{O}^{+}$ and $\mathrm{O}^{2+}$ ionic abundances. Points are colour-coded according to the ionisation
parameter, $\log U_*$, derived by {\sc HCm}.}

   \label{comp-icf}
\end{figure}

To explore more directly the origin of the systematic difference in S/H, we compared the sulphur ICF implied by the AGN photoionisation models with the prescription derived by \cite{dors16} from photoionisation models of \hii\ regions. The latter can be written as

\begin{equation}
{\rm ICF}_{\rm D16}(S)=
\left[
1-\left(1-\frac{O^+}{O^++O^{2+}}\right)^\alpha
\right]^{-1/\alpha},
\end{equation}

\noindent with \(\alpha=3.27\). For each object in the control sample, we calculated this ICF using its measured \(O^+\) and \(O^{2+}\) abundances. The corresponding {\sc HCm} value was obtained directly from the total and ionic-equivalent sulphur abundances predicted by the models contributing to the {\sc HCm} solution,

Figure \ref{comp-icf} shows a clear systematic difference between the two prescriptions. All 45 objects have \({\rm ICF}_{\rm HCm}>{\rm ICF}_{\rm D16}\), with median values of 3.48 and 1.01, respectively. The median difference, expressed as \(\log({\rm ICF}_{\rm HCm}/{\rm ICF}_{\rm D16})\), is 0.51 dex. Moreover, this difference increases strongly with the ionisation parameter (\(\rho=0.87\), \(p=7.2\times10^{-15}\), Spearman test). This behaviour indicates that an ICF calibrated from \hii\ region photoionisation models does not adequately account for the higher ionisation stages of sulphur under the harder ionising conditions characteristic of AGN, with the discrepancy becoming particularly important at high \(U\). This provides a direct explanation for the systematically higher total S/H abundances obtained with {\sc HCm}.

\subsection{Application to MaNGA AGN}\label{sec:manga}

\begin{figure*}
   \centering
   \includegraphics[width=0.85\textwidth,clip=]{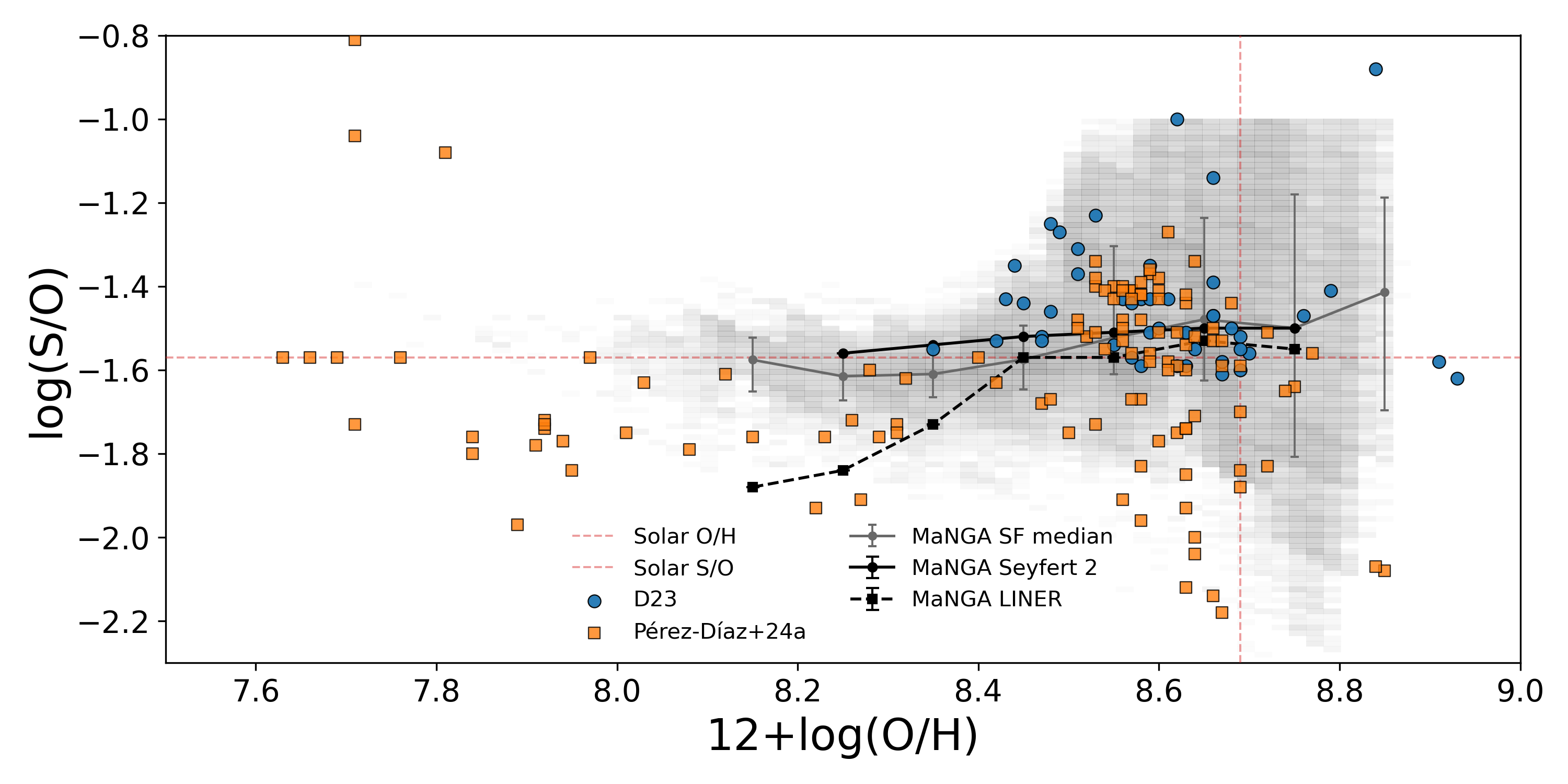}
\caption{Relation between sulphur-to-oxygen abundance ratio, $\log(\mathrm{S/O})$, and oxygen abundance, $12+\log(\mathrm{O/H})$, for different classes of ionised regions and AGN samples. The grey density map shows the number of MaNGA star-forming regions per bin on a logarithmic scale from \cite{pm25b} analysed with {\sc HCm}, while the grey solid line shows their binned median relation. Black solid circles and line correspond to the median values for MaNGA Seyfert~2 regions, while black dashed squares and line show the median trend for MaNGA LINER regions; error bars indicate the dispersion within each abundance bin. Blue circles represent the nuclear AGN sample from D23 analysed with {\sc HCm}, while orange squares correspond to the sample from \cite{pd24a} analysed with {\sc HCm-IR}. Horizontal and vertical red dashed lines mark the solar sulphur-to-oxygen abundance ratio and solar oxygen abundance, respectively.}

   \label{OH-SO}
\end{figure*}

\begin{figure*}
   \centering
   \includegraphics[width=0.85\textwidth,clip=]{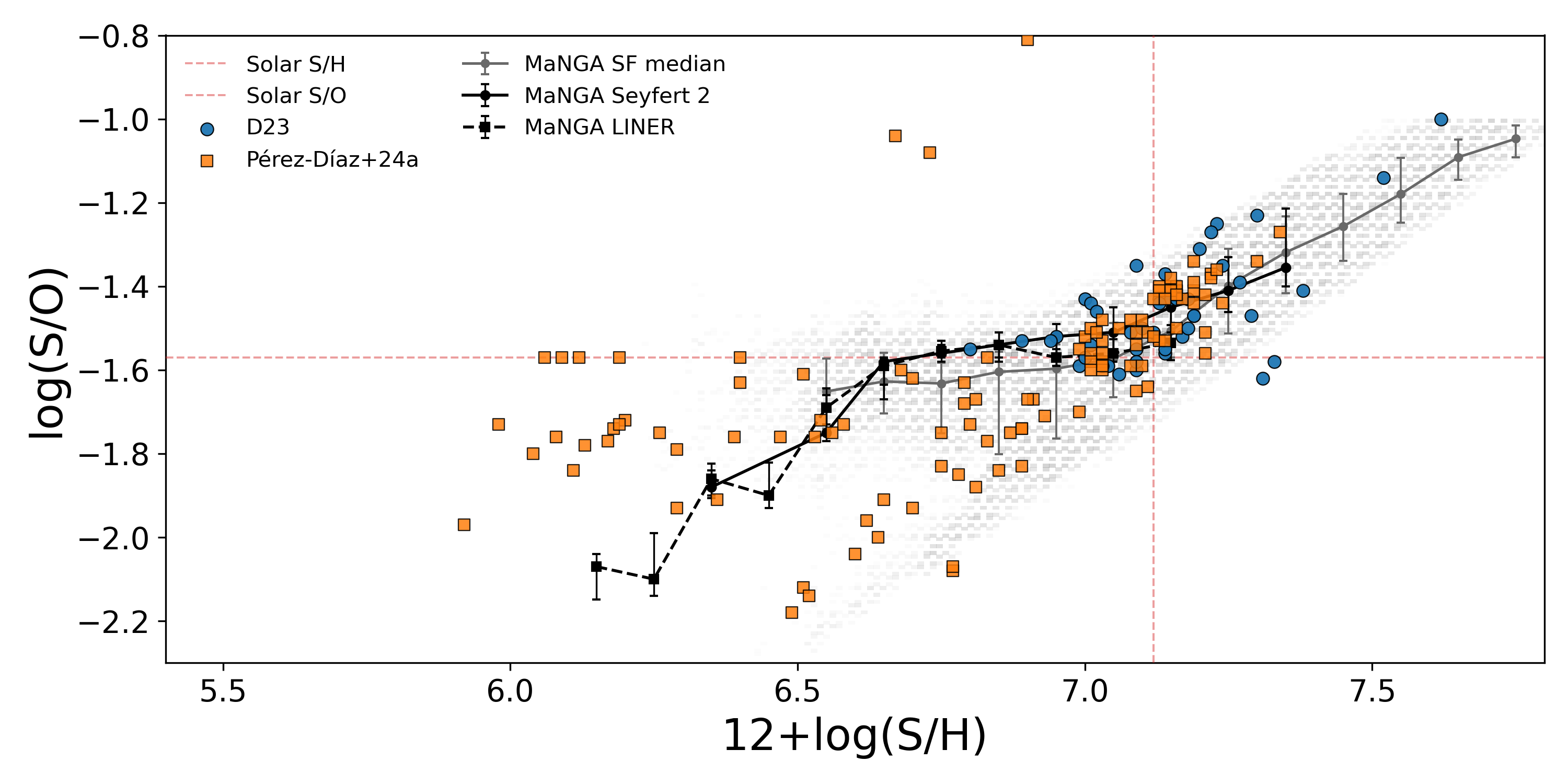}
\caption{
Relation between sulphur-to-oxygen abundance ratio, $\log(\mathrm{S/O})$, and sulphur abundance, $12+\log(\mathrm{S/H})$, for the same samples shown in Fig. \ref{OH-SO}. Symbols and colour coding are identical to those in Fig. \ref{OH-SO}. The grey density map shows the number of MaNGA star-forming regions per bin on a logarithmic scale  from \cite{pm25b}, while the grey solid line shows its binned median relation. Horizontal and vertical red dashed lines indicate the solar $\mathrm{S/O}$ ratio and solar sulphur abundance, respectively. 
}

   \label{SH-SO}
\end{figure*}

The distributions of O/H and S/H allow us to test whether Seyfert and LINER-like regions occupy different abundance regimes. The S/O distribution is particularly informative because it is sensitive to the relative behaviour of oxygen and sulphur, to the adopted ionisation correction implicit in the models, and potentially to depletion effects in dusty nuclear environments.
To do so, we applied \textsc{HCm} to the MaNGA AGN-like sample to derive O/H, S/H, and $\log U_*$ and  we analysed the results separately for Seyfert and LINER-like regions, considering for all objects the same model conditions, as fixed by consistency with the direct method in the previous subsection for the control sample, that is, a double-peaked power-law with $\alpha_{\textrm ox}$ = -1.2 and a stopping criterion with a $f_e$ = 98\%. 
The uncertainties in the abundances derived for the MaNGA samples were obtained by propagating the observational uncertainties of the emission-line fluxes through the Monte Carlo procedure implemented in {\sc HCm}. They therefore represent observational uncertainties and do not include systematic effects associated with the adopted model grid.
In the next subsection, different SEDs will be discussed in the case of LINERs.

However, prior to this calculation, it is convenient to examine possible external contributions to the most relevant emission lines used in the analysis.

For instance, as discussed in \cite{pm23} for MaNGA data, the [\sii] emission, fundamental for the calculation of S/H abundances in the optical,  can be contaminated by a non-negligible excess coming from DIG. This excess was later quantified in the case of star-forming pointings by \cite{pm25b} by comparing the N/O estimation as obtained by {\sc HCm} when calculated using the lines involved in the N2O2 and N2S2 parameters. It was observed by \cite{z21} that N/O abundances are systematically lower in the latter case, so this was interpreted as an excess of the [\sii] emission. In the case of star-forming pointings this excess also shows a clear anti-correlation with EW(\ha), so its natural DIG origin is well justified.

Nonetheless, in the case of our AGN sample, a Spearman rank test reveals only weak correlations between the [\sii] excess and EW(\ha), with opposite signs for the two AGN classes: \(\rho=-0.135\) (\(p=4.5\times10^{-11}\)) for Seyfert 2 galaxies and \(\rho=0.226\) (\(p=2.2\times10^{-6}\)) for LINERs. Although formally significant owing to the large sample sizes, particularly for Seyfert 2 galaxies, the small correlation coefficients and the opposite directions of the trends do not indicate a common systematic dependence of the [\sii] excess on EW(\ha) that could be attributed to DIG contamination, so we decided not to apply any correction in this sample.
“This lack of a clear DIG-like dependence of the [\sii] excess on EW(\ha) is consistent with the expectation that DIG contamination does not significantly contribute to the central kpc regions, where the bulk of the AGN activity is located, and further supports the AGN nature of the ionising source in our sample, including LINERs.

Figures \ref{OH-SO} and \ref{SH-SO} compare the sulphur-to-oxygen abundance ratio derived with {\sc HCm} for spatially resolved star-forming regions and AGN-ionised environments using two different abundance planes: $\log(\mathrm{S/O})$ versus $12+\log(\mathrm{O/H})$ and $\log(\mathrm{S/O})$ versus $12+\log(\mathrm{S/H})$. Together, these diagrams provide insight into whether the behaviour observed in AGN environments reflects genuine abundance differences or instead arises from ionisation effects, dust depletion, or methodological systematics.

The large distribution of $\sim$ 200\,000 MaNGA \hii\ regions analysed in \cite{pm25b} defines a well-populated reference sequence. In the oxygen-based representation Fig. \ref{OH-SO}, the star-forming population occupies a relatively narrow locus concentrated around approximately solar metallicity, with most regions distributed between $12+\log(\mathrm{O/H})\sim8.3$--$8.8$ and $\log(\mathrm{S/O})\sim-1.9$ to $-1.4$. The density distribution reveals a modest tendency toward increasing sulphur-to-oxygen ratio at high oxygen abundance, consistent with the behaviour already discussed in \cite{pm25b}.

When the Seyfert and LINER median relations are compared with the star-forming sequence, a more nuanced picture emerges. Over most of the abundance range, the median S/O ratios of the three populations remain relatively similar. At the highest oxygen abundances, Seyfert regions still tend to display slightly larger median S/O values than star-forming regions, although the difference is modest and the two populations remain largely consistent within the observed dispersion. The Seyfert distribution also presents a substantially larger scatter, extending towards higher S/O values.

This tendency is also reflected by the nuclear AGN sample from D23, several of whose objects populate the upper envelope of the Seyfert distribution and display among the highest S/O ratios. Part of this small offset may reflect intrinsic differences between both samples. The D23 abundances were derived from integrated nuclear spectra obtained through single-aperture observations, whereas the MaNGA measurements correspond to spatially resolved regions extracted from integral-field spectroscopy, covering a wider area contaminated with the ISM surrounding the nuclear regions. As already suggested by the comparison in the BPT diagram (Section 2), the two datasets do not sample exactly the same physical scales or nuclear environments. Consequently, a modest systematic difference in their derived abundances is not unexpected, although the overall agreement between both samples remains remarkably good, as also seen in Fig. \ref{s2-s3}.

In contrast, the LINER median sequence remains much closer to that of the star-forming regions over the full metallicity range.
The sample from \cite{pd24a} exhibits substantially larger scatter than both the MaNGA Seyfert and the samples from D23. Unlike the other datasets shown here, the abundance determinations for this sample in \cite{pd24a} were obtained using infrared emission lines within {\sc HCm-IR}.  In that work, the oxygen abundance was not derived from oxygen emission lines, but from neon lines under the assumption of a solar Ne/O ratio. Consequently, the inferred O/H values are largely independent of oxygen depletion onto dust grains, allowing a cleaner comparison between sulphur and oxygen abundances than in optical determinations based directly on oxygen emission lines. Furthermore, this sample contains both star-forming and AGN-ionised systems, increasing the intrinsic diversity of physical conditions represented. The fact that the infrared-based sample does not exhibit the same modest differences observed in the optical nuclear AGN samples may therefore provide an additional indication that depletion effects contribute significantly to the sulphur-to-oxygen behaviour inferred from optical diagnostics.

Despite these differences, all samples preserve the same overall tendency for S/O to increase with metallicity. This behaviour becomes much clearer when sulphur abundance is adopted as metallicity tracer (Fig. \ref{SH-SO}), where all three populations define similar increasing sequences in the S/O-S/H plane. The modest differences observed in the oxygen-based representation becomes substantially weaker, and the median relations are largely consistent within their corresponding dispersions over most of the abundance range. Nevertheless, Seyfert regions still tend to occupy the upper envelope of the relation, particularly at the highest sulphur abundances.
This comparison suggests that a significant fraction of the separation observed in the O/H-S/O plane arises from the choice of oxygen as a metallicity tracer rather than from a genuine systematic enhancement of sulphur. However, the residual tendency of Seyfert regions to define the upper envelope of the S/O-S/H relation indicates that additional physical effects may therefore contribute.
In fact, the use of sulphur abundance as a metallicity tracer provides an important physical advantage. Unlike oxygen, sulphur is expected to be much less affected by depletion onto dust grains and therefore may provide a cleaner tracer of the total gas-phase metallicity. Although sulphur formally still appears in the definition of $\log(\mathrm{S/O})$, the much tighter sequence observed in the $12+\log(\mathrm{S/H})$ plane suggests that sulphur abundance better isolates abundance effects from depletion-related processes.

However, this interpretation is not unique, and the comparison should be treated with caution, since the samples shown in Figs. \ref{OH-SO} and \ref{SH-SO} are heterogeneous in terms of observational wavelength range, aperture, abundance methodology, and sample selection.
In any case, the persistence of this small residual displacement suggests that depletion or projection effects alone may not fully explain the observed behaviour. Some additional physical mechanism may therefore contribute to the tendency towards elevated sulphur-to-oxygen ratios observed in part of the AGN-ionised population.
The comparison between MaNGA Seyfert regions and the D23 sample provides additional insight. This latter sample corresponds to integrated nuclear spectra, whereas the MaNGA measurements trace individual spatially resolved regions. Nuclear integrated spectra are therefore expected to sample physical conditions weighted toward the innermost narrow-line region, potentially characterised by larger dust content, stronger depletion effects, or AGN-specific ionisation conditions. Aperture effects may therefore naturally explain why the D23 sample occupies the upper envelope of the Seyfert distribution.

Beyond depletion and ionisation effects, galaxy-scale processes may also play a role. Although more speculative, galaxy-scale gas flows associated with AGN activity may also contribute to the observed behaviour. The central regions of AGN host galaxies are complex environments where inflows, outflows, and feedback-driven gas motions may modify the local chemical properties of the interstellar medium. Metal transport induced by AGN feedback may either enhance or dilute the abundances measured in the nuclear regions depending on the balance between centrally concentrated enrichment and gas mixing processes.
Such a scenario may be related to galaxy-scale gas flows associated with AGN activity, as suggested by spatially resolved studies reporting evidence for metal-poor gas accretion and a diversity of metallicity gradients in AGN-ionised gas \citep{doNascimento2022, Amiri2024, Armah2024}.
This possibility becomes particularly relevant when comparing the different samples analysed here. The spatially resolved MaNGA regions probe localised ionised structures, whereas the D23 sample traces integrated nuclear spectra that are more directly sensitive to the central metal reservoir. Differences in the maximum metallicities reached by Seyfert nuclei, LINER regions, and star-forming environments may therefore partly encode information about metal redistribution processes operating in AGN host galaxies.

The comparison between the oxygen-based and sulphur-based abundance planes provides an opportunity to investigate this possibility. Since sulphur abundance may provide a more robust tracer of the total gas-phase metallicity, modest differences in the high-metallicity regime between AGN populations could help disentangle depletion effects from genuine metal redistribution processes associated with AGN feedback.
Future work combining optical and infrared diagnostics within a unified {\sc HCm} framework will be essential to disentangle these contributions and determine whether different AGN populations require distinct ionising prescriptions to reproduce their observed abundances.

\begin{figure*}
   \centering
   \includegraphics[width=0.9\textwidth,clip=]{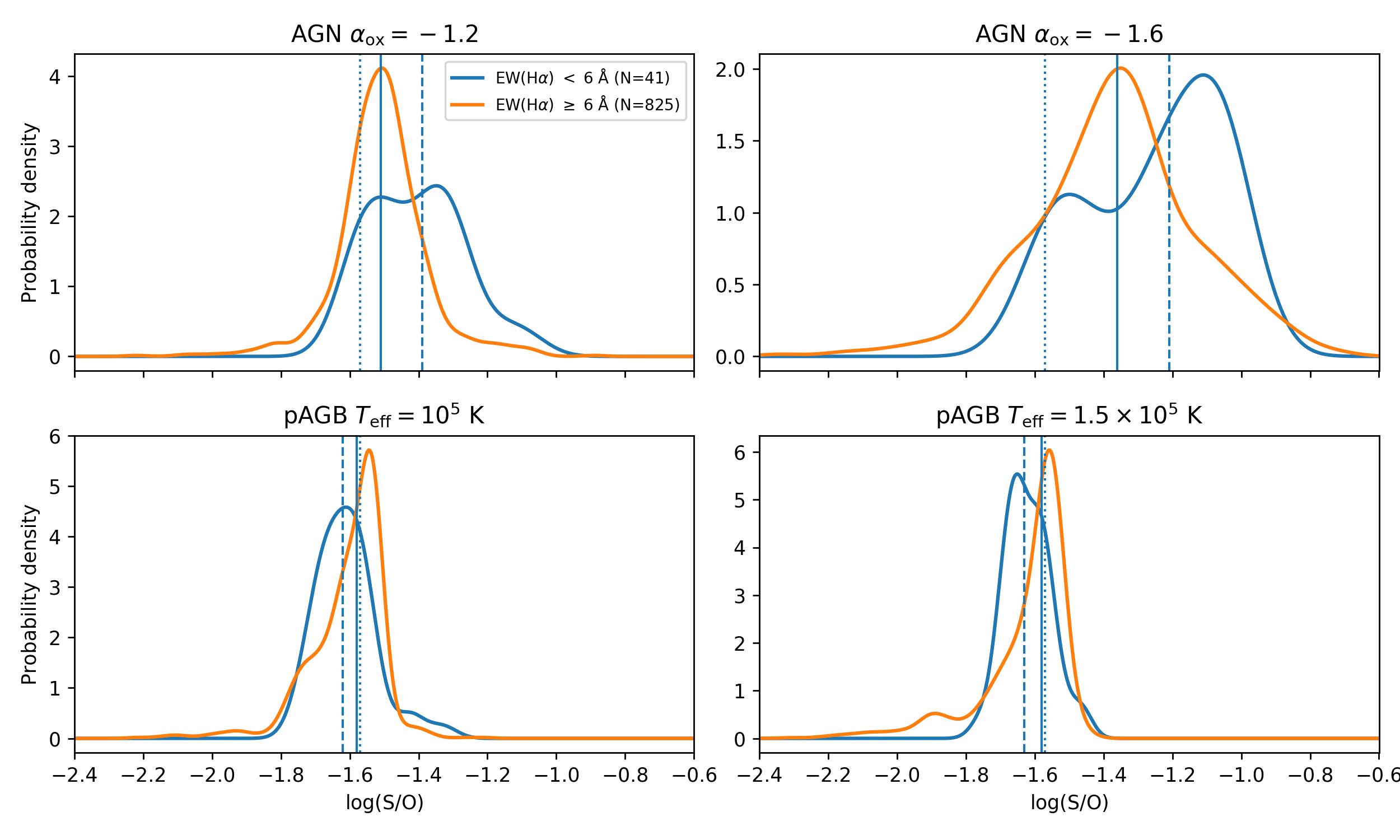}

\caption{
Distribution of sulphur-to-oxygen abundance ratios derived for the MaNGA LINER sample using different ionising SEDs within {\sc HCm}. The four panels correspond to AGN photoionisation models with $\alpha_{\rm ox}=-1.2$ and $-1.6$, and pAGB stellar populations with effective temperatures of $10^{5}$ and $1.5\times10^{5}$ K.
Orange and blue curves show the kernel density distributions for regions with EW(\ha) $\geq$ 6 \AA\ and EW(\ha) $<$ 6 \AA, respectively, normalised independently to account for the different sizes of the two subsamples.
Solid and dashed vertical lines indicate the median values of the high- and low-EW subsamples, respectively, while the dotted line marks the solar sulphur-to-oxygen abundance ratio. 
}

   \label{SO_liner}
\end{figure*}

\subsection{The ionising SED of LINERs}

The interpretation of the LINER population remains one of the most debated issues in studies of low-ionisation emission-line regions. While classical Seyfert nuclei are widely accepted as being powered by accretion onto supermassive black holes, the nature of LINER emission is still under discussion, with proposed ionising sources including low-luminosity AGN, pAGB stars, and combinations of stellar and AGN-related ionisation mechanisms.

In the previous section, we showed that assuming the AGN model grid favoured by the control sample, namely $\alpha_{\rm ox}=-1.2$ and a free-electron fraction of 98\%, LINER regions tend to display lower sulphur-to-oxygen abundance ratios than Seyfert 2 regions, while remaining generally close to the star-forming sequence. In this sense, the abundance sequence traced by the different AGN samples can be qualitatively summarised as follows: MaNGA LINERs generally show lower S/O ratios than MaNGA Seyfert 2 regions, while the highest values are reached by some objects in the nuclear Seyfert sample from D23.
The fact that LINERs occupy an intermediate position is particularly interesting because all MaNGA AGN abundances presented in this work were derived using exactly the same {\sc HCm} model grid. Therefore, the differences observed between Seyfert and LINER populations cannot be attributed to different abundance methodologies or model assumptions. Instead, they must reflect either genuine differences in the physical properties of the ionised gas or limitations of the adopted ionising continuum in describing the LINER population.

To investigate this possibility, we recalculated the LINER sample using alternative ionising SEDs. Figure \ref{SO_liner} shows the resulting distributions of S/O obtained for AGN models with $\alpha_{\rm ox}=-1.2$ and $-1.6$, together with pAGB stellar populations characterised by effective temperatures of $10^{5}$ and $1.5\times10^{5}$ K. The derived sulphur-to-oxygen abundance ratios depend significantly on the adopted ionising continuum. Softer AGN continua produce systematically larger S/O values, whereas pAGB models yield narrower distributions concentrated around values close to the solar ratio.

The comparison with \ha\ EWs provides additional information. The selected LINER sample contains only a small fraction of regions with EW(\ha) $<$ 6 \AA, with 41 objects compared to 825 regions above this threshold. 
However, this fraction should be interpreted with caution, since the S/N \(\geq 10\) requirement imposed on the [\siii] lines may preferentially exclude low-EW(\ha), low-surface-brightness LINER regions. Therefore, the EW(\ha) distribution of our LINER sample is affected by the selection function, and the small fraction of objects below 6 \AA\ cannot by itself exclude a larger contribution from pAGB ionisation in the parent LINER population.
The kernel density distributions show that the differences between the low- and high-EW subsamples depend on the adopted ionising continuum. For the pAGB models, both subsamples display relatively similar S/O distributions, concentrated around values close to the solar ratio, whereas the AGN models produce larger differences in both the location and width of the distributions, with the low-EW subsample generally extending towards higher S/O values. Given the limited number of objects with EW(\ha) $<$ 6 \AA, however, the detailed shape of its distribution should not be overinterpreted.
This behaviour contrasts with the results obtained by \cite{pd25}, where a substantial fraction of the LINER population occupied the regime traditionally associated with retired galaxies in the WHAN diagram. The scarcity of low-EW objects in the present sample therefore suggests that the LINER regions analysed here are not representative of the full LINER population identified in MaNGA, but instead correspond preferentially to regions with stronger ionised-gas emission.
This result weakens the interpretation of the present sample as being dominated by pAGB ionisation. If the observed LINER emission were primarily powered by evolved stellar populations, a larger fraction of low-EW objects would be expected. Instead, the abundance of regions with EW(H$\alpha$)$>6$ \AA\ points toward a population more closely related to genuine AGN activity.

The comparison with the results of \cite{pd25} is particularly informative. In that work, {\sc HCm} was applied to MaNGA LINERs using several families of ionising continua, including AGN, ADAF and pAGB models. The derived oxygen abundances were found to depend significantly on the adopted ionising source, whereas N/O remained comparatively robust. Furthermore, the infrared analysis presented in \cite{pm25a} suggested the existence of two preferred AGN populations characterised by different values of $\alpha_{\rm ox}$, with one family favouring systematically lower values than the more accepted $\alpha_{\rm ox}=-1.2$ solution.
An additional aspect highlighted by the comparison with \cite{pd25} is the diagnostic potential of the S/O ratio itself. The behaviour found here suggests that S/O may provide an additional and complementary constraint on the ionisation structure.
 Unlike O/H, which primarily traces the global metallicity of the gas, the sulphur-to-oxygen ratio appears to be particularly sensitive to the relative distribution of ionic stages predicted by different photoionisation models. The systematic differences observed between Seyfert and LINER populations therefore reinforce the idea that S/O may constitute a valuable diagnostic of the underlying ionising source and its SED.

At first sight, the present S/O analysis might appear difficult to reconcile with that picture. The softer AGN models tested here produce larger sulphur-to-oxygen ratios, whereas the observed LINER population exhibits systematically lower S/O values than Seyfert~2 nuclei. However, this apparent contradiction arises only if the derived S/O values are interpreted as a direct measure of spectral hardness. The behaviour of Figure \ref{SO_liner} instead demonstrates that the inferred S/O ratio depends on the adopted ionisation structure and on the relative populations of the different sulphur ions. Consequently, changes in the recovered sulphur abundance do not necessarily translate into a straightforward measure of the hardness of the ionising continuum.
For this reason, the preferred interpretation of the present results is not that LINERs require softer continua than Seyfert nuclei. Rather, assuming that the control sample correctly identifies $\alpha_{\rm ox}=-1.2$ as the most realistic AGN ionising continuum, the lower S/O values derived for LINERs indicate that these objects occupy an intermediate regime between star-forming regions and classical Seyfert nuclei. Whether this difference reflects distinct SEDs, different ionisation parameters, variations in gas geometry, or differences in dust depletion remains an open question.

Ultimately, a definitive interpretation requires extending the present analysis to additional ionising continua and testing whether alternative LINER-specific model grids provide a better description of the observed abundances. In particular, a control sample analogous to that used for Seyfert nuclei, but composed of LINERs with direct-method ionic abundances, would provide a crucial benchmark for determining whether the differences observed here arise from genuine physical distinctions or from limitations in the adopted photoionisation models.

\section{Summary and conclusions}

In this work, we have extended the \textsc{HCm} methodology to derive total sulphur abundances in AGN from optical emission lines through comparison with photoionisation models, avoiding the use of empirical sulphur ionisation correction factors. The comparison with a control sample of Seyfert galaxies shows that the code successfully reproduces both oxygen and sulphur  ionic abundances derived through the direct method when AGN models with $\alpha_{\rm ox}=-1.2$ and a stopping criterion corresponding to a free-electron fraction of 98\% are adopted. 
The fact that the same model configuration simultaneously reproduces the ionic abundance scale and adequately covers the observed sulphur diagnostic space provides an independent consistency check of the adopted photoionisation grid.

However, although ionic abundances are accurately reproduced, the total sulphur abundances derived by \textsc{HCm} are systematically larger than those obtained from direct-method analyses using classical sulphur ionisation correction factors. This result strongly suggests that ICF prescriptions calibrated for \hii\ regions underestimate the contribution of higher sulphur ionisation stages under AGN ionisation conditions. Consequently, sulphur abundances in AGN may have been systematically underestimated when derived using standard optical ICF schemes.

Applying the method to a large MaNGA AGN sample reveals that Seyfert regions tend to present larger median S/O ratios than star-forming regions only at the highest oxygen abundances, whereas LINERs remain much closer to the star-forming sequence. This difference is accompanied by a substantially larger dispersion among Seyfert regions.

Both the O/H-S/O and the S/H-S/O diagrams reveal the same positive correlation between S/O and metallicity. Adopting sulphur as an alternative metallicity tracer does not alter this global chemical trend, but substantially reduces the apparent offset between star-forming regions and Seyferts, indicating that part of the separation observed in the oxygen-based diagram is linked to the choice of the metallicity tracer.
Nevertheless, Seyfert regions still tend to populate the upper envelope of the relation at the highest abundances, leaving room for additional effects beyond oxygen depletion.

The remaining differences are consistent with the combined action of AGN-specific ionisation structure, aperture-dependent sampling between integrated nuclear and spatially resolved observations, and possibly intrinsic differences in the ionising continuum. Additional galaxy-scale gas flows associated with AGN activity cannot be excluded.
The origin of the differences observed between Seyfert and LINER regions remains uncertain. 
Since both populations were analysed using the same family of AGN photoionisation models, the observed offsets are unlikely to arise solely from differences in the analysis methodology.
They may instead reflect differences in ionisation structure, depletion, gas geometry, or in the shape of the ionising SED. In particular, the lower S/O ratios derived for LINERs relative to Seyfert nuclei suggest differences in their ionisation conditions, although their abundance sequence remains much closer to that of star-forming regions.

The adopted AGN model configuration was calibrated using the Seyfert control sample, for which the agreement between ionic abundances derived through the direct method and those predicted by \textsc{HCm} is very good. However, an equivalent control sample is presently unavailable for LINERs. Future analyses of LINERs with direct-method ionic abundances will therefore be required to establish whether the differences observed here arise from genuine physical distinctions or instead reveal the need for alternative ionising continua or geometrical prescriptions. In this context, recent infrared studies have suggested the existence of different preferred values of $\alpha_{\rm ox}$ among AGN populations, although the connection between those infrared populations and the classical optical Seyfert--LINER classification remains unclear.

\begin{acknowledgments}
This work has been funded by project Estallidos8 PID2022-136598NB-C32  (Spanish Ministerio de Ciencia e Innovaci\'on). We also acknowledge financial support from the Severo Ochoa grant CEX2021-001131-S funded by MCIN/AEI/10.13039/501100011033. EPM also acknowledges the assistance from his guide dog Rocko, without whose daily help this work would have been much more difficult.
IAZ acknowledges funding from the Deutsche Forschungsgemeinschaft (DFG; German Research Foundation)---project-ID 550945879.
\end{acknowledgments}


\bibliographystyle{aa}
\typeout{}
\bibliography{HCm-AGN-sulphur_OJAp}

\begin{thebibliography}{55}
\expandafter\ifx\csname natexlab\endcsname\relax\def\natexlab#1{#1}\fi

\bibitem[{{Abdurro'uf} {et~al.}(2022){Abdurro'uf}, {Accetta}, {Aerts}, {Silva
  Aguirre}, {Ahumada}, {Ajgaonkar}, {Filiz Ak}, {Alam}, {Allende Prieto},
  {Almeida}, {Anders}, {Anderson}, {Andrews}, {Anguiano}, {Aquino-Ort{\'\i}z},
  {Arag{\'o}n-Salamanca}, {Argudo-Fern{\'a}ndez}, {Ata}, {Aubert},
  {Avila-Reese}, {Badenes}, {Barb{\'a}}, {Barger}, {Barrera-Ballesteros},
  {Beaton}, {Beers}, {Belfiore}, {Bender}, {Bernardi}, {Bershady}, {Beutler},
  {Bidin}, {Bird}, {Bizyaev}, {Blanc}, {Blanton}, {Boardman}, {Bolton},
  {Boquien}, {Borissova}, {Bovy}, {Brandt}, {Brown}, {Brownstein}, {Brusa},
  {Buchner}, {Bundy}, {Burchett}, {Bureau}, {Burgasser}, {Cabang}, {Campbell},
  {Cappellari}, {Carlberg}, {Wanderley}, {Carrera}, {Cash}, {Chen}, {Chen},
  {Cherinka}, {Chiappini}, {Choi}, {Chojnowski}, {Chung}, {Clerc}, {Cohen},
  {Comerford}, {Comparat}, {da Costa}, {Covey}, {Crane}, {Cruz-Gonzalez},
  {Culhane}, {Cunha}, {Dai}, {Damke}, {Darling}, {Davidson}, {Davies},
  {Dawson}, {De Lee}, {Diamond-Stanic}, {Cano-D{\'\i}az}, {S{\'a}nchez},
  {Donor}, {Duckworth}, {Dwelly}, {Eisenstein}, {Elsworth}, {Emsellem},
  {Eracleous}, {Escoffier}, {Fan}, {Farr}, {Feng}, {Fern{\'a}ndez-Trincado},
  {Feuillet}, {Filipp}, {Fillingham}, {Frinchaboy}, {Fromenteau}, {Galbany},
  {Garc{\'\i}a}, {Garc{\'\i}a-Hern{\'a}ndez}, {Ge}, {Geisler}, {Gelfand},
  {G{\'e}ron}, {Gibson}, {Goddy}, {Godoy-Rivera}, {Grabowski}, {Green},
  {Greener}, {Grier}, {Griffith}, {Guo}, {Guy}, {Hadjara}, {Harding},
  {Hasselquist}, {Hayes}, {Hearty}, {Hern{\'a}ndez}, {Hill}, {Hogg},
  {Holtzman}, {Horta}, {Hsieh}, {Hsu}, {Hsu}, {Huber}, {Huertas-Company},
  {Hutchinson}, {Hwang}, {Ibarra-Medel}, {Chitham}, {Ilha}, {Imig}, {Jaekle},
  {Jayasinghe}, {Ji}, {Johnson}, {Jones}, {J{\"o}nsson}, {Katkov}, {Khalatyan},
  {Kinemuchi}, {Kisku}, {Knapen}, {Kneib}, {Kollmeier}, {Kong}, {Kounkel},
  {Kreckel}, {Krishnarao}, {Lacerna}, {Lane}, {Langgin}, {Lavender}, {Law},
  {Lazarz}, {Leung}, {Leung}, {Lewis}, {Li}, {Li}, {Lian}, {Liang}, {Lin},
  {Lin}, {Lin}, {Lintott}, {Long}, {Longa-Pe{\~n}a}, {L{\'o}pez-Cob{\'a}},
  {Lu}, {Lundgren}, {Luo}, {Mackereth}, {de la Macorra}, {Mahadevan},
  {Majewski}, {Manchado}, {Mandeville}, {Maraston}, {Margalef-Bentabol},
  {Masseron}, {Masters}, {Mathur}, {McDermid}, {Mckay}, {Merloni},
  {Merrifield}, {Meszaros}, {Miglio}, {Di Mille}, {Minniti}, {Minsley},
  {Monachesi}, {Moon}, {Mosser}, {Mulchaey}, {Muna}, {Mu{\~n}oz}, {Myers},
  {Myers}, {Nadathur}, {Nair}, {Nandra}, {Neumann}, {Newman}, {Nidever},
  {Nikakhtar}, {Nitschelm}, {O'Connell}, {Garma-Oehmichen}, {Luan Souza de
  Oliveira}, {Olney}, {Oravetz}, {Ortigoza-Urdaneta}, {Osorio}, {Otter},
  {Pace}, {Padilla}, {Pan}, {Pan}, {Parikh}, {Parker}, {Peirani}, {Pe{\~n}a
  Ram{\'\i}rez}, {Penny}, {Percival}, {Perez-Fournon}, {Pinsonneault},
  {Poidevin}, {Poovelil}, {Price-Whelan}, {B{\'a}rbara de Andrade Queiroz},
  {Raddick}, {Ray}, {Rembold}, {Riddle}, {Riffel}, {Riffel}, {Rix}, {Robin},
  {Rodr{\'\i}guez-Puebla}, {Roman-Lopes}, {Rom{\'a}n-Z{\'u}{\~n}iga}, {Rose},
  {Ross}, {Rossi}, {Rubin}, {Salvato}, {S{\'a}nchez}, {S{\'a}nchez-Gallego},
  {Sanderson}, {Santana Rojas}, {Sarceno}, {Sarmiento}, {Sayres}, {Sazonova},
  {Schaefer}, {Schiavon}, {Schlegel}, {Schneider}, {Schultheis}, {Schwope},
  {Serenelli}, {Serna}, {Shao}, {Shapiro}, {Sharma}, {Shen}, {Shetrone}, {Shu},
  {Simon}, {Skrutskie}, {Smethurst}, {Smith}, {Sobeck}, {Spoo}, {Sprague},
  {Stark}, {Stassun}, {Steinmetz}, {Stello}, {Stone-Martinez},
  {Storchi-Bergmann}, {Stringfellow}, {Stutz}, {Su}, {Taghizadeh-Popp},
  {Talbot}, {Tayar}, {Telles}, {Teske}, {Thakar}, {Theissen}, {Tkachenko},
  {Thomas}, {Tojeiro}, {Hernandez Toledo}, {Troup}, {Trump}, {Trussler},
  {Turner}, {Tuttle}, {Unda-Sanzana}, {V{\'a}zquez-Mata}, {Valentini},
  {Valenzuela}, {Vargas-Gonz{\'a}lez}, {Vargas-Maga{\~n}a}, {Alfaro},
  {Villanova}, {Vincenzo}, {Wake}, {Warfield}, {Washington}, {Weaver},
  {Weijmans}, {Weinberg}, {Weiss}, {Westfall}, {Wild}, {Wilde}, {Wilson},
  {Wilson}, {Wilson}, {Wolf}, {Wood-Vasey}, {Yan}, {Zamora}, {Zasowski},
  {Zhang}, {Zhao}, {Zheng}, {Zheng}, \& {Zhu}}]{sdss-dr17}
{Abdurro'uf}, {Accetta}, K., {Aerts}, C., {et~al.} 2022, \apjs, 259, 35

\bibitem[{{Amiri} {et~al.}(2024){Amiri}, {Knapen}, {Comer{\'o}n}, {Marconi}, \&
  {Lehmer}}]{Amiri2024}
{Amiri}, A., {Knapen}, J.~H., {Comer{\'o}n}, S., {Marconi}, A., \& {Lehmer},
  B.~D. 2024, \aap, 689, A193

\bibitem[{{Armah} {et~al.}(2024){Armah}, {Riffel}, {Dahmer-Hahn}, {Davies},
  {Dors}, {Kakkad}, {Riffel}, {Rodr{\'\i}guez-Ardila}, {Ruschel-Dutra}, \&
  {Storchi-Bergmann}}]{Armah2024}
{Armah}, M., {Riffel}, R., {Dahmer-Hahn}, L.~G., {et~al.} 2024, \mnras, 534,
  2723

\bibitem[{{Asari} {et~al.}(2007){Asari}, {Cid Fernandes}, {Stasi{\'n}ska},
  {Torres-Papaqui}, {Mateus}, {Sodr{\'e}}, {Schoenell}, \& {Gomes}}]{Asari2007}
{Asari}, N.~V., {Cid Fernandes}, R., {Stasi{\'n}ska}, G., {et~al.} 2007,
  \mnras, 381, 263

\bibitem[{{Asplund} {et~al.}(2009){Asplund}, {Grevesse}, {Sauval}, \&
  {Scott}}]{asplund}
{Asplund}, M., {Grevesse}, N., {Sauval}, A.~J., \& {Scott}, P. 2009, \araa, 47,
  481

\bibitem[{{Baldwin} {et~al.}(1981){Baldwin}, {Phillips}, \& {Terlevich}}]{bpt}
{Baldwin}, J.~A., {Phillips}, M.~M., \& {Terlevich}, R. 1981, \pasp, 93, 5

\bibitem[{{Binette} {et~al.}(2022){Binette}, {Villar Mart{\'\i}n}, {Magris C.},
  {Mart{\'\i}nez-Paredes}, {Alarie}, {Rodr{\'\i}guez Ardila}, \&
  {Villica{\~n}a-Pedraza}}]{binette22}
{Binette}, L., {Villar Mart{\'\i}n}, M., {Magris C.}, G., {et~al.} 2022,
  \rmxaa, 58, 133

\bibitem[{{Blanton} {et~al.}(2017){Blanton}, {Bershady}, {Abolfathi},
  {Albareti}, {Allende Prieto}, {Almeida}, {Alonso-Garc{\'{\i}}a}, {Anders},
  {Anderson}, {Andrews}, \& et~al.}]{Blanton2017}
{Blanton}, M.~R., {Bershady}, M.~A., {Abolfathi}, B., {et~al.} 2017, \aj, 154,
  28

\bibitem[{{Bundy} {et~al.}(2015){Bundy}, {Bershady}, {Law}, {Yan}, {Drory},
  {MacDonald}, {Wake}, {Cherinka}, {S{\'a}nchez-Gallego}, {Weijmans}, {Thomas},
  {Tremonti}, {Masters}, {Coccato}, {Diamond-Stanic}, {Arag{\'o}n-Salamanca},
  {Avila-Reese}, {Badenes}, {Falc{\'o}n-Barroso}, {Belfiore}, {Bizyaev},
  {Blanc}, {Bland-Hawthorn}, {Blanton}, {Brownstein}, {Byler}, {Cappellari},
  {Conroy}, {Dutton}, {Emsellem}, {Etherington}, {Frinchaboy}, {Fu}, {Gunn},
  {Harding}, {Johnston}, {Kauffmann}, {Kinemuchi}, {Klaene}, {Knapen},
  {Leauthaud}, {Li}, {Lin}, {Maiolino}, {Malanushenko}, {Malanushenko}, {Mao},
  {Maraston}, {McDermid}, {Merrifield}, {Nichol}, {Oravetz}, {Pan}, {Parejko},
  {Sanchez}, {Schlegel}, {Simmons}, {Steele}, {Steinmetz}, {Thanjavur},
  {Thompson}, {Tinker}, {van den Bosch}, {Westfall}, {Wilkinson}, {Wright},
  {Xiao}, \& {Zhang}}]{Bundy2015}
{Bundy}, K., {Bershady}, M.~A., {Law}, D.~R., {et~al.} 2015, \apj, 798, 7

\bibitem[{{Cid Fernandes} {et~al.}(2005){Cid Fernandes}, {Mateus}, {Sodr{\'e}},
  {Stasi{\'n}ska}, \& {Gomes}}]{CidFernandes2005}
{Cid Fernandes}, R., {Mateus}, A., {Sodr{\'e}}, L., {Stasi{\'n}ska}, G., \&
  {Gomes}, J.~M. 2005, \mnras, 358, 363

\bibitem[{{Cid Fernandes} {et~al.}(2010){Cid Fernandes}, {Stasi{\'n}ska},
  {Schlickmann}, {Mateus}, {Vale Asari}, {Schoenell}, \& {Sodr{\'e}}}]{cid10}
{Cid Fernandes}, R., {Stasi{\'n}ska}, G., {Schlickmann}, M.~S., {et~al.} 2010,
  \mnras, 403, 1036

\bibitem[{{D{\'{\i}}az}(1998)}]{diaz98}
{D{\'{\i}}az}, {\'A}.~I. 1998, \apss, 263, 143

\bibitem[{{D{\'\i}az} \& {P{\'e}rez-Montero}(2000)}]{dpm00}
{D{\'\i}az}, A.~I. \& {P{\'e}rez-Montero}, E. 2000, \mnras, 312, 130

\bibitem[{{do Nascimento} {et~al.}(2022){do Nascimento}, {Dors},
  {Storchi-Bergmann}, {Mallmann}, {Riffel}, {Ilha}, {Riffel}, {Rembold},
  {Deconto-Machado}, {da Costa}, \& {Armah}}]{doNascimento2022}
{do Nascimento}, J.~C., {Dors}, O.~L., {Storchi-Bergmann}, T., {et~al.} 2022,
  \mnras, 513, 807

\bibitem[{{Dors} {et~al.}(2014){Dors}, {Cardaci}, {H{\"a}gele}, \&
  {Krabbe}}]{dors14}
{Dors}, O.~L., {Cardaci}, M.~V., {H{\"a}gele}, G.~F., \& {Krabbe}, {\^A}.~C.
  2014, \mnras, 443, 1291

\bibitem[{{Dors} {et~al.}(2021){Dors}, {Contini}, {Riffel},
  {P{\'e}rez-Montero}, {Krabbe}, {Cardaci}, \& {H{\"a}gele}}]{dors21}
{Dors}, O.~L., {Contini}, M., {Riffel}, R.~A., {et~al.} 2021, \mnras, 501, 1370

\bibitem[{{Dors} {et~al.}(2016){Dors}, {P{\'e}rez-Montero}, {H{\"a}gele},
  {Cardaci}, \& {Krabbe}}]{dors16}
{Dors}, O.~L., {P{\'e}rez-Montero}, E., {H{\"a}gele}, G.~F., {Cardaci}, M.~V.,
  \& {Krabbe}, A.~C. 2016, \mnras, 456, 4407

\bibitem[{{Dors} {et~al.}(2023){Dors}, {Valerdi}, {Riffel}, {Riffel},
  {Cardaci}, {H{\"a}gele}, {Armah}, {Revalski}, {Flury}, {Freitas-Lemes},
  {Am{\^o}res}, {Krabbe}, {Binette}, {Feltre}, \& {Storchi-Bergmann}}]{dors23}
{Dors}, O.~L., {Valerdi}, M., {Riffel}, R.~A., {et~al.} 2023, \mnras, 521, 1969

\bibitem[{{Ferland} {et~al.}(2017){Ferland}, {Chatzikos}, {Guzm{\'a}n},
  {Lykins}, {van Hoof}, {Williams}, {Abel}, {Badnell}, {Keenan}, {Porter}, \&
  {Stancil}}]{cloudy}
{Ferland}, G.~J., {Chatzikos}, M., {Guzm{\'a}n}, F., {et~al.} 2017, \rmxaa, 53,
  385

\bibitem[{{Fern{\'a}ndez-Ontiveros} {et~al.}(2021){Fern{\'a}ndez-Ontiveros},
  {P{\'e}rez-Montero}, {V{\'\i}lchez}, {Amor{\'\i}n}, \& {Spinoglio}}]{jafo21b}
{Fern{\'a}ndez-Ontiveros}, J.~A., {P{\'e}rez-Montero}, E., {V{\'\i}lchez},
  J.~M., {Amor{\'\i}n}, R., \& {Spinoglio}, L. 2021, \aap, 652, A23

\bibitem[{{Flury} \& {Moran}(2020)}]{flury20}
{Flury}, S.~R. \& {Moran}, E.~C. 2020, \mnras, 496, 2191

\bibitem[{{Garnett}(1989)}]{garnett89}
{Garnett}, D.~R. 1989, \apj, 345, 282

\bibitem[{{Garnett} {et~al.}(1997){Garnett}, {Shields}, {Skillman}, {Sagan}, \&
  {Dufour}}]{garnett97}
{Garnett}, D.~R., {Shields}, G.~A., {Skillman}, E.~D., {Sagan}, S.~P., \&
  {Dufour}, R.~J. 1997, \apj, 489, 63

\bibitem[{{Izotov} {et~al.}(1994){Izotov}, {Thuan}, \&
  {Lipovetsky}}]{Izotov1994}
{Izotov}, Y.~I., {Thuan}, T.~X., \& {Lipovetsky}, V.~A. 1994, \apj, 435, 647

\bibitem[{{Kauffmann} {et~al.}(2003){Kauffmann}, {Heckman}, {Tremonti},
  {Brinchmann}, {Charlot}, {White}, {Ridgway}, {Brinkmann}, {Fukugita}, {Hall},
  {Ivezi{\'c}}, {Richards}, \& {Schneider}}]{kauffman03}
{Kauffmann}, G., {Heckman}, T.~M., {Tremonti}, C., {et~al.} 2003, \mnras, 346,
  1055

\bibitem[{{Kehrig} {et~al.}(2006){Kehrig}, {V{\'\i}lchez}, {Telles},
  {Cuisinier}, \& {P{\'e}rez-Montero}}]{kehrig06}
{Kehrig}, C., {V{\'\i}lchez}, J.~M., {Telles}, E., {Cuisinier}, F., \&
  {P{\'e}rez-Montero}, E. 2006, \aap, 457, 477

\bibitem[{{Kewley} {et~al.}(2001){Kewley}, {Dopita}, {Sutherland}, {Heisler},
  \& {Trevena}}]{kewley01}
{Kewley}, L.~J., {Dopita}, M.~A., {Sutherland}, R.~S., {Heisler}, C.~A., \&
  {Trevena}, J. 2001, \apj, 556, 121

\bibitem[{{Kewley} {et~al.}(2006){Kewley}, {Groves}, {Kauffmann}, \&
  {Heckman}}]{kewley06}
{Kewley}, L.~J., {Groves}, B., {Kauffmann}, G., \& {Heckman}, T. 2006, \mnras,
  372, 961

\bibitem[{{Luridiana} {et~al.}(2015){Luridiana}, {Morisset}, \& {Shaw}}]{pyneb}
{Luridiana}, V., {Morisset}, C., \& {Shaw}, R.~A. 2015, \aap, 573, A42

\bibitem[{{Mateus} {et~al.}(2006){Mateus}, {Sodr{\'e}}, {Cid Fernandes},
  {Stasi{\'n}ska}, {Schoenell}, \& {Gomes}}]{Mateus2006}
{Mateus}, A., {Sodr{\'e}}, L., {Cid Fernandes}, R., {et~al.} 2006, \mnras, 370,
  721

\bibitem[{{Monteiro} \& {Dors}(2021)}]{monteiro21}
{Monteiro}, A.~F. \& {Dors}, O.~L. 2021, \mnras, 508, 3023

\bibitem[{{Nagao} {et~al.}(2006){Nagao}, {Maiolino}, \& {Marconi}}]{nagao06}
{Nagao}, T., {Maiolino}, R., \& {Marconi}, A. 2006, \aap, 447, 863

\bibitem[{{Nemmen} {et~al.}(2014){Nemmen}, {Storchi-Bergmann}, \&
  {Eracleous}}]{adaf}
{Nemmen}, R.~S., {Storchi-Bergmann}, T., \& {Eracleous}, M. 2014, \mnras, 438,
  2804

\bibitem[{{Peimbert} \& {Costero}(1969)}]{pc69}
{Peimbert}, M. \& {Costero}, R. 1969, Boletin de los Observatorios Tonantzintla
  y Tacubaya, 5, 3

\bibitem[{{P{\'e}rez-D{\'\i}az} {et~al.}(2022){P{\'e}rez-D{\'\i}az},
  {P{\'e}rez-Montero}, {Fern{\'a}ndez-Ontiveros}, \& {V{\'\i}lchez}}]{pd22}
{P{\'e}rez-D{\'\i}az}, B., {P{\'e}rez-Montero}, E., {Fern{\'a}ndez-Ontiveros},
  J.~A., \& {V{\'\i}lchez}, J.~M. 2022, \aap, 666, A115

\bibitem[{{P{\'e}rez-D{\'\i}az} {et~al.}(2024){P{\'e}rez-D{\'\i}az},
  {P{\'e}rez-Montero}, {Fern{\'a}ndez-Ontiveros}, {V{\'\i}lchez},
  {Hern{\'a}n-Caballero}, \& {Amor{\'\i}n}}]{pd24a}
{P{\'e}rez-D{\'\i}az}, B., {P{\'e}rez-Montero}, E., {Fern{\'a}ndez-Ontiveros},
  J.~A., {et~al.} 2024, \aap, 685, A168

\bibitem[{{P{\'e}rez-D{\'\i}az} {et~al.}(2025){P{\'e}rez-D{\'\i}az},
  {P{\'e}rez-Montero}, {Zinchenko}, \& {V{\'\i}lchez}}]{pd25}
{P{\'e}rez-D{\'\i}az}, B., {P{\'e}rez-Montero}, E., {Zinchenko}, I.~A., \&
  {V{\'\i}lchez}, J.~M. 2025, \aap, 694, A18

\bibitem[{{P{\'e}rez-Montero}(2014)}]{hcm14}
{P{\'e}rez-Montero}, E. 2014, \mnras, 441, 2663

\bibitem[{{P{\'e}rez-Montero} \& {D{\'\i}az}(2005)}]{pmd05}
{P{\'e}rez-Montero}, E. \& {D{\'\i}az}, A.~I. 2005, \mnras, 361, 1063

\bibitem[{{P{\'e}rez-Montero} {et~al.}(2006){P{\'e}rez-Montero}, {D{\'\i}az},
  {V{\'\i}lchez}, \& {Kehrig}}]{pm06}
{P{\'e}rez-Montero}, E., {D{\'\i}az}, A.~I., {V{\'\i}lchez}, J.~M., \&
  {Kehrig}, C. 2006, \aap, 449, 193

\bibitem[{{P{\'e}rez-Montero} {et~al.}(2019){P{\'e}rez-Montero}, {Dors},
  {V{\'\i}lchez}, {Garc{\'\i}a-Benito}, {Cardaci}, \& {H{\"a}gele}}]{hcm-agn}
{P{\'e}rez-Montero}, E., {Dors}, O.~L., {V{\'\i}lchez}, J.~M., {et~al.} 2019,
  \mnras, 489, 2652

\bibitem[{{P{\'e}rez-Montero} {et~al.}(2025{\natexlab{a}}){P{\'e}rez-Montero},
  {Fern{\'a}ndez-Ontiveros}, {P{\'e}rez-D{\'\i}az}, {V{\'\i}lchez}, \&
  {Amor{\'\i}n}}]{pm25a}
{P{\'e}rez-Montero}, E., {Fern{\'a}ndez-Ontiveros}, J.~A.,
  {P{\'e}rez-D{\'\i}az}, B., {V{\'\i}lchez}, J.~M., \& {Amor{\'\i}n}, R.
  2025{\natexlab{a}}, \aap, 696, A229

\bibitem[{{P{\'e}rez-Montero} {et~al.}(2025{\natexlab{b}}){P{\'e}rez-Montero},
  {P{\'e}rez-D{\'\i}az}, {V{\'\i}lchez}, {Zinchenko}, {Castrillo},
  {Gavil{\'a}n}, {Zamora}, \& {D{\'\i}az}}]{pm25b}
{P{\'e}rez-Montero}, E., {P{\'e}rez-D{\'\i}az}, B., {V{\'\i}lchez}, J.~M.,
  {et~al.} 2025{\natexlab{b}}, The Open Journal of Astrophysics, 8, 51253

\bibitem[{{P{\'e}rez-Montero} {et~al.}(2023){P{\'e}rez-Montero}, {Zinchenko},
  {V{\'\i}lchez}, {Zurita}, {Florido}, \& {P{\'e}rez-D{\'\i}az}}]{pm23}
{P{\'e}rez-Montero}, E., {Zinchenko}, I.~A., {V{\'\i}lchez}, J.~M., {et~al.}
  2023, \aap, 669, A88

\bibitem[{{Rauch}(2003)}]{rauch}
{Rauch}, T. 2003, \aap, 403, 709

\bibitem[{{Riffel} {et~al.}(2006){Riffel}, {Rodr{\'\i}guez-Ardila}, \&
  {Pastoriza}}]{riffel06}
{Riffel}, R., {Rodr{\'\i}guez-Ardila}, A., \& {Pastoriza}, M.~G. 2006, \aap,
  457, 61

\bibitem[{{Rodr{\'\i}guez-Ardila} {et~al.}(2017){Rodr{\'\i}guez-Ardila},
  {Mason}, {Martins}, {Ramos Almeida}, {Riffel}, {Riffel}, {Lira},
  {Gonz{\'a}lez Mart{\'\i}n}, {Dametto}, {Flohic}, {Ho}, {Ruschel-Dutra},
  {Thanjavur}, {Colina}, {McDermid}, {Perlman}, \& {Winge}}]{Rodriguez2017}
{Rodr{\'\i}guez-Ardila}, A., {Mason}, R.~E., {Martins}, L., {et~al.} 2017,
  \mnras, 465, 906

\bibitem[{{Savage} \& {Sembach}(1996)}]{savage96}
{Savage}, B.~D. \& {Sembach}, K.~R. 1996, \araa, 34, 279

\bibitem[{{Stasi{\'n}ska}(1978)}]{stasinska78}
{Stasi{\'n}ska}, G. 1978, \aap, 66, 257

\bibitem[{{Vermeij} \& {van der Hulst}(2002)}]{vermeij02}
{Vermeij}, R. \& {van der Hulst}, J.~M. 2002, \aap, 391, 1081

\bibitem[{{Vilchez} \& {Esteban}(1996)}]{ve96}
{Vilchez}, J.~M. \& {Esteban}, C. 1996, \mnras, 280, 720

\bibitem[{{Zhang} {et~al.}(2013){Zhang}, {Liang}, \& {Hammer}}]{zhang13}
{Zhang}, Z.~T., {Liang}, Y.~C., \& {Hammer}, F. 2013, \mnras, 430, 2605

\bibitem[{{Zhu} {et~al.}(2024){Zhu}, {Kewley}, \& {Sutherland}}]{zhu24}
{Zhu}, P., {Kewley}, L.~J., \& {Sutherland}, R.~S. 2024, \apj, 977, 187

\bibitem[{{Zinchenko} {et~al.}(2016){Zinchenko}, {Pilyugin}, {Grebel},
  {S{\'a}nchez}, \& {V{\'{\i}}lchez}}]{zinchenko2016}
{Zinchenko}, I.~A., {Pilyugin}, L.~S., {Grebel}, E.~K., {S{\'a}nchez}, S.~F.,
  \& {V{\'{\i}}lchez}, J.~M. 2016, \mnras, 462, 2715

\bibitem[{{Zinchenko} {et~al.}(2021){Zinchenko}, {V{\'\i}lchez},
  {P{\'e}rez-Montero}, {Sukhorukov}, {Sobolenko}, \& {Duarte Puertas}}]{z21}
{Zinchenko}, I.~A., {V{\'\i}lchez}, J.~M., {P{\'e}rez-Montero}, E., {et~al.}
  2021, \aap, 655, A58

\end{thebibliography}


\end{document}